\documentclass[journal,twocolumn]{IEEEtranTCOM}
\normalsize

\ifCLASSINFOpdf
\else
\fi

\ifCLASSINFOpdf
\usepackage[pdftex]{graphicx}
\else
\usepackage[dvips]{graphicx}
\fi
\usepackage{url}

\usepackage[cmex10]{amsmath}
\makeatletter
\g@addto@macro\normalsize{%
	\setlength\abovedisplayskip{2pt}
	\setlength\belowdisplayskip{2pt}
	\setlength\abovedisplayshortskip{2pt}
	\setlength\belowdisplayshortskip{2pt}
}
\makeatother
\usepackage{amssymb}
\usepackage{color}

\allowdisplaybreaks

\usepackage{array}

\usepackage{cite}
\usepackage{bbm}

\newtheorem{theorem}{Theorem}
\newtheorem{definition}{Definition}

\newtheorem{remark}{Remark}
\newcommand{\GW}{\textnormal{GW}}
\newcommand{\SI}{\textnormal{SI}}

\usepackage[text={6.5in,9.5in},centering]{geometry}

\usepackage{setspace}
\begin{document}

%
\title{On Hypothesis Testing Against Conditional Independence with Multiple Decision Centers}

\author{Sadaf Salehkalaibar, Mich\`ele Wigger, and Roy Timo
		\thanks{S.~Salehkalaibar is with the Department of Electrical and Computer Engineering, College of Engineering, University of Tehran, Tehran, Iran, s.saleh@ut.ac.ir,}
	\thanks{
		M.~Wigger is with   LTCI,  Telecom ParisTech, Universit\'e Paris-Saclay, 75013 Paris, michele.wigger@telecom-paristech.fr,}

	\thanks{R.~Timo is with Ericsson Research, Stockholm, Sweden, roy.timo@ericsson.com, }
	\thanks{The work of M. Wigger was supported by the ERC Grant through CTO Com. Parts of the material in this paper have been presented at IEEE SPCOM Systems (ISWCS), Bangalore, India, June 2016.}
	
}

\maketitle

\begin{abstract}  A distributed binary hypothesis testing problem is studied  with one observer and two decision centers. Achievable  type-II error exponents are derived for testing against conditional independence when  the observer communicates with the two decision centers over one common and two individual noise-free bit pipes and when it communicates with them over a noisy broadcast channel (BC).  The results are based on a coding and testing scheme that splits the observations into subblocks, so that transmitter and receivers can independently apply to each subblock either  Gray-Wyner coordination coding with side-information or hybrid joint source-channel coding with side-information, followed by a Neyman-Pearson test over the subblocks at the receivers. This approach allows to avoid introducing further error exponents that one would expect from the receivers' decoding operations related to   binning or the noisy transmission channel. 
The derived exponents are shown to be optimal in some special cases when communication is over noise-free links. The results  reveal a tradeoff between the type-II error exponents at the two decision centers.

\end{abstract}

\IEEEpeerreviewmaketitle

\section{Introduction}

Consider the distributed hypothesis testing problem where a transmitter  communicates with two receivers that each wishes to decide on the joint probability distribution underlying the observations 
at the three terminals. 
In the scenario we consider, communication from the transmitter to the receivers either takes place over one common and two individual noise-free bit pipes or over a discrete memoryless broadcast channel (BC). 
For simplicity, we restrict attention to a \emph{binary} hypothesis where either $\mathcal{H}=0$ or $\mathcal{H}=1$.  The focus  of this paper is on the asymptotic regime where the length of the observed sequences $n$ tends to infinity and where  both the type-I error probabilities (i.e., the probabilities of deciding on hypothesis $1$ when $\mathcal{H}=0$) and the type-II error probabilities (i.e., the probabilities of deciding on hypothesis $0$ when $\mathcal{H}=1$)  vanish. We follow the approach in \cite{Ahlswede,Han}, and aim to quantify  the fastest possible exponential decrease of the type-II error probabilities, while we allow the type-I error probabilities to vanish arbitrarily slowly. Ahlswede and Csiszar \cite{Ahlswede} and Han \cite{Han} studied the problem with only a single receiver and where communication takes place over a noise-free link. They presented general upper and lower bounds on the maximum type-II error exponents, and these bounds match when under $\mathcal{H}=1$ the joint distribution of the observations $X^n$ at the transmitter and $Y^n$ at the receiver equals the product of the marginal distributions under $\mathcal{H}=0$. This problem formulation is widely known as \emph{testing against independence}.
Rahman and Wagner \cite{Wagner} extended this result to a  setup called \emph{testing against conditional independence} where the receiver observes two sequences $(Y^n, Z^n)$:  under both hypotheses, sequence $Z^n$ has  the same joint distribution with the transmitter's observation $X^n$ and the same joint distribution with $Y^n$; and under $\mathcal{H}=1$, observation $Y^n$ is conditionally independent of $X^n$ given $Z^n$. Similar results were also found for scenarios with multiple transmitters \cite{Han, Wagner}, interactive transmitters, interactive multi-round communications between nodes, successive refinement and privacy setups \cite{Kim, Debbah, Lai2, Tan}. \setlength{\belowcaptionskip}{-10pt}
\begin{figure}[t]
	\centering
	\includegraphics[scale=0.32,angle=0]{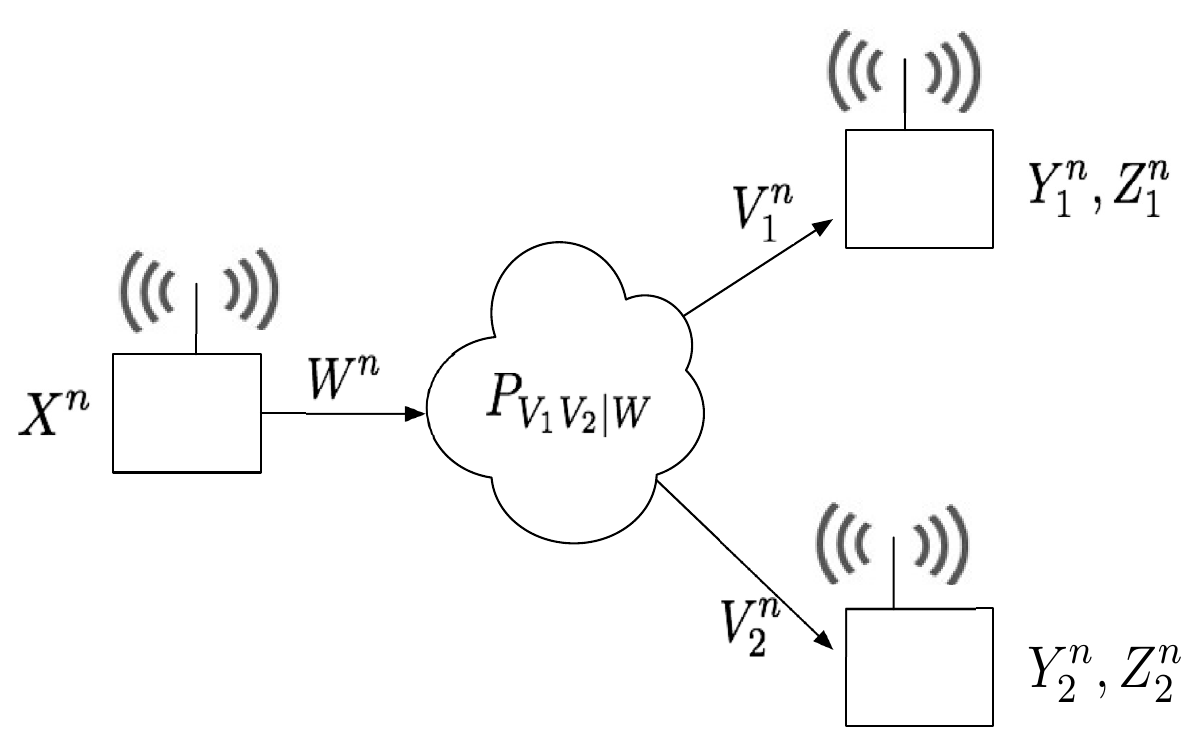}

	\caption{Multi-terminal hypothesis testing with side information.}
	\vspace{0cm}
	\label{Figure 3}
\end{figure}

 When testing against conditional independence, in contrast to the simpler testing against independence, a code construction with binning \cite{Amari,Wagner} has to be used to send information from the transmitter to the receiver. The roles of the two receiver observations $Z^n$ and $Y^n$ decouple:  $Z^n$ plays the role of side-information for the source-coding scheme and thus reduces the required communication rate by means of binning;  
 $Y^n$ is solely used for hypothesis testing but not for recovering the correct codeword. 
 Generally, the decoding operation at the receiver introduced by binning causes  a second competing error exponent compared to the standard scheme where the codeword index is directly sent over the channel \cite{Amari}. In the special case of testing against conditional independence, the second error exponent  is however inactive. Rahman and Wagner \cite{Wagner} proposed a multi-letter extension of the binning scheme and an analysis of this scheme that directly proves the final result with the single error exponent.  

A similar technique was recently applied also by Sreekuma and Gunduz \cite{Gunduz} to derive the optimal error exponent for testing against conditional independence over a discrete memoryless  channel (DMC). Their result shows that in this special case, the same error exponent can be achieved as when communication is over a noise-free link of rate equal to the capacity of the DMC. Surprisingly, there is thus no competing error exponent caused by the noisy communication channel. The work in \cite{Gunduz} also extends some of the results to a scenario with multiple transmitters.

In contrast to these previous  works, here we  consider a single transmitter and \emph{multiple receivers with different local observations}.  The goal is to understand the tension on the communication channel caused by the receivers being interested in learning different informations from the transmitter. 

Multiple receivers with different observations  can   be used to model a variety of situations: 
\begin{itemize}
	\item \textit{Multiple Decision Centers Deciding on Different Hypotheses:} Multiple decision centers wish to decide on the same binary hypothesis but they have different local informations. This work treats the scenario where communication to the decision centers takes place over a common network. 
	
\underline{Example 1:}  Consider a road-side sensor which measures road conditions  (e.g., wetness) and vehicles parameters (e.g., speed or inter-car distances). Suppose that there are two autonomous cars which measure the same parameters using the on-board sensors. Each of them verifies the accuracy of its own measurements by comparing its data to the data collected at the road-side sensors: if the sets of data are independent, then the car decides  that its own data is faulty and raises an alarm (or goes to a predefined mode).\\

	\item \textit{Single Decision Center with Uncertain Local Observation:} There is only a single decision center,  and  the probability distribution of the decision center's observation under each of the two hypotheses is unknown to the transmitter. In this case, the transmitter has to code for both options simultaneously, and our results determine the exponent pairs that are simultaneously achievable for the two options.

	\underline{Example 2:}  Consider an earthquake alert system with a remote sensor and a single local decision center that also senses ground vibrations. At unknown times of the day, there is heavy traffic close to the decision center and thus the sensed vibrations follow a different distribution. In this scenario, the information communicated from the sensor to the decision center needs to be useful under both traffic conditions. Testing against (conditional) independence can be used to distinguish vibrations that are independent at the sensor and the decision center and thus not coming from larger-scale seismic activities.
	\\
	
		\item \textit{Single Decision Center Performing Two Simultaneous Tests:} Assume there is a single decision center with two sets of observations $(Y_1^n, Z_1^n)$ and $(Y_2^n, Z_2^n)$ that wishes to decide on two hypotheses and it suffices to take each decision only based on one of the two sets of observations. For example, because $(Y_2^n, Z_2^n)$ is irrelevant for the first hypothesis test given $(Y_1^n, Z_1^n)$ and the opposite holds for the second hypothesis test.

	\underline{Example 3:} Consider a remote combined  temperature and humidity sensor  and a local weather station that also senses these two phenomena but can well separate the two measurements. For simplicity, the local station might then choose to  decide on the temperature to forecast based only on its temperature measurement and  to predict  the humidity only based on the humidity measurement.
	\\
\end{itemize}
A main feature of the scenario that we consider is that the observer is interested in extracting and transmitting information about its observation $X^n$ that is useful to both receivers. There is thus an inherent tradeoff in the problem, in that some information might be more beneficial for Receiver~1 than for Receiver~2 and vice versa. 
 The goal of this paper is to shed light on this tradeoff when testing against conditional independence.  As will be explained shortly, we consider communications of positive rates. Interestingly, for zero-rate communication, such a tradeoff never exists. That means, there is a single strategy at the transmitter that is optimal for both decision centers. This optimal strategy  is simply the strategy from  \cite{Amari, Han} where the transmitter sends a single bit indicating whether its observation is typical with respect to the distribution under $\mathcal{H}=0$, irrespective of the distribution of the receiver observation.

One of the main contributions of this paper is to propose and analyze a coding and testing scheme for testing against conditional independence with two receivers either over a source coding network with a common and two individual noise-free bit-pipes or over a discrete memoryless BC. In both scenarios, there is a single type-II error exponent as in the scenario with a single receiver. Moreover, the  decoding operations  at the receivers only limit the rate of communication and the bin sizes that one is allowed to choose, but do not introduce a second competing error exponent.  In our scheme, each terminal splits its observation into many subblocks and then applies either a Gray-Wyner coordination coding  scheme  with side-information \cite{Gray,Shayevitz} or a hybrid source-channel coding scheme \cite{Minero} to each subblock, and each receiver performs a Neyman-Pearson test over all these subblocks to decide on the desired hypothesis. The idea of using block coding followed by a Neyman-Pearson test is inspired by \cite{Wagner} and \cite{Gunduz}. However, here we use different block codings compared to the works in \cite{Wagner} and \cite{Gunduz}, as these latter only consider only a single decision center. Moreover, we perform the Neyman-Pearson  test over the reconstructed codeword sequences and not directly over the transmitted messages or channel outputs. This approach allows to simplify the analysis compared to an analysis that closely follows the steps proposed in \cite{Wagner} for the single-decision center scenario. 

The second main contribution of the paper is to show that the proposed schemes achieve the optimal type-II error exponents when \emph{testing against independence} over a common and two individual noise-free bit-pipes and when \emph{testing against conditional independence} only over a common pipe under some less-noisy assumptions on the side-informations. For this latter result, a Gaussian example is presented that clearly illustrates the tradeoff on the communication channel  stemming from the presence of two decision centers.

%

%
%
%
%
%

\subsection{Notation}\label{sec2}

Random variables are denoted by capital letters, e.g., $X$, $Y$, and their realizations by lower case letters, e.g., $x$, $y$.  Script symbols  such as $\mathcal{X}$ and $\mathcal{Y}$ stand for alphabets of  random variables and realizations, and $\mathcal{X}^n$ and $\mathcal{Y}^n$ for the corresponding $n$-fold Cartesian products.  Sequences of random variables $(X_i,...,X_j)$ and realizations $(x_i,\ldots, x_j)$ are  abbreviated by $X_i^j$ and $x_{i}^j$.  When $i=1$, then we also use the notations $X^j$ and $x^j$ instead of $X_1^j$ and $x_{1}^j$. 

The probability mass function (pmf) of a finite random variable $X$ is written as $P_X$; the conditional pmf of $X$ given $Y$ is written as $P_{X|Y}$. Entropy, conditional entropy, and mutual information of random variables $X$ and $Y$  are  denoted by $H(X)$, $H(X|Y)$, and $I(X;Y)$. Differential entropy and conditional differential entropy of continuous random variables $X$ and $Y$ are  indicated by $h(X)$ and $h(X|Y)$. All entropies  and mutual informations in this paper are meant with respect to the distribution under hypothesis $\mathcal{H}=0$. The term   $D(P||Q)$ stands for  the Kullback-Leibler divergence between two pmfs $P$ and $Q$ over the same alphabet.  

For a given $P_X$ and a constant $\mu>0$, let   $\mathcal{T}_{\mu}^n(P_X)=\{x^n:| \# \{i:x_i=x\}/n-P_X(x)|\leq \mu P_X(x), \forall x\in\mathcal{X}\}$ be the set of $\mu$-typical sequences in $\mathcal{X}^n$ \cite{ElGamal}. Similarly,   $\mathcal{T}_{\mu}^n(P_{X,Y})$ stands for the set of jointly $\mu$-typical sequences.

The expectation operator is written as $\mathbb{E}[.]$.  A Gaussian distribution with mean $a$ and variance $\sigma^2$ is written as $\mathcal{N}(a, \sigma^2)$. We  abbreviate \emph{independent and identically distributed} by \emph{i.i.d.}.
Finally, the $\log(.)$-function is  taken with respect to base 2.

\section{Hypothesis Testing Over a Gray-Wyner Netwrok with Side Information}\label{newsec}
  \begin{figure*}[t]
	\centering
	\includegraphics[scale=0.3,angle=0]{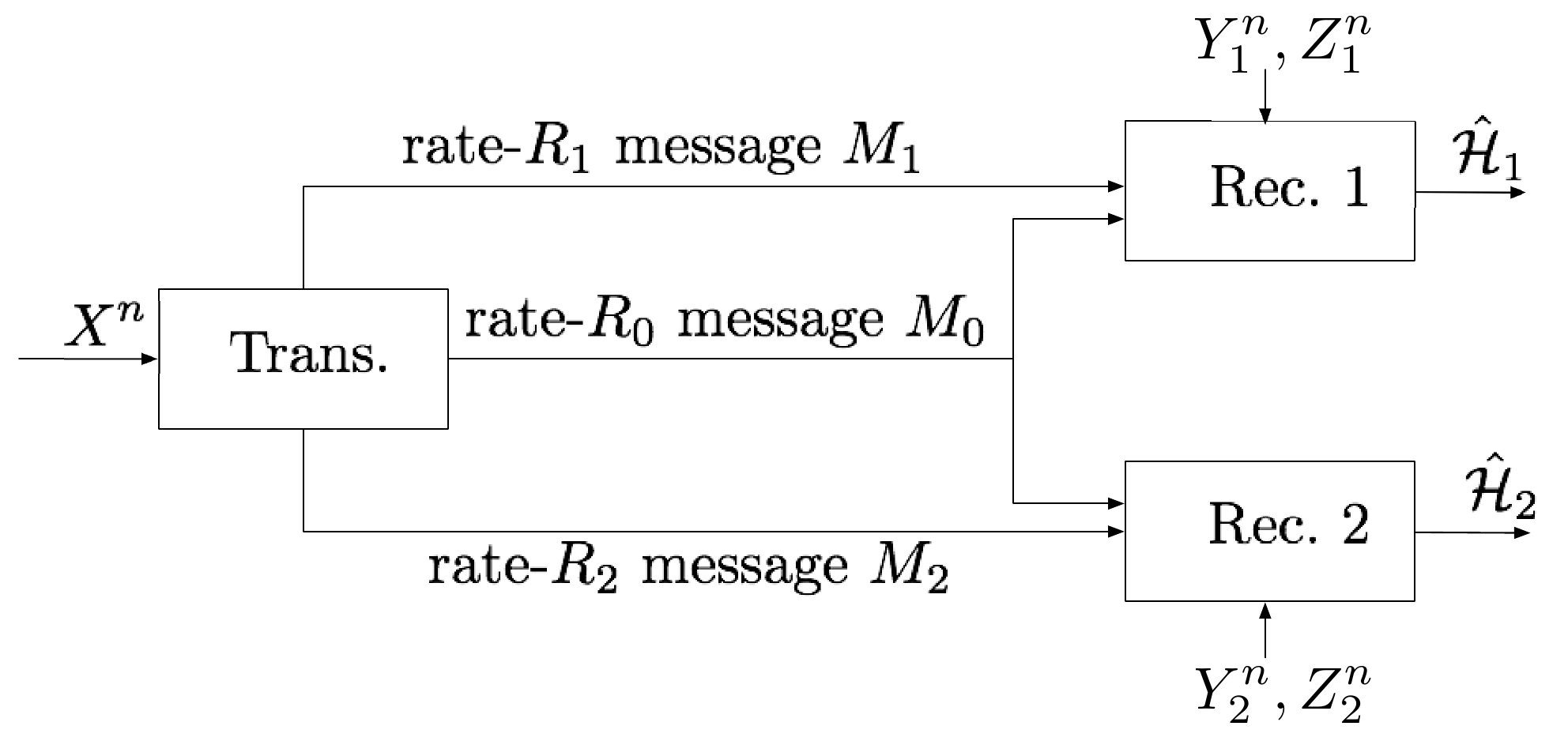}
	\vspace{-0.5cm}
	\caption{Hypothesis testing over a Gray-Wyner network with side information.}
	\vspace{-0.5cm}
	\label{Figure new}
\end{figure*}

Consider the distributed hypothesis testing problem with one transmitter  and two receivers in Fig.~\ref{Figure new}. The transmitter observes the sequence $X^n$, and Receivers~1 and 2 observe $Y_1^n$ and $Y_2^n$, respectively. In this model, for $i\in\{1,2\}$, Receiver~$i$   additionally also observes a side information $Z_i^n$ whose 
\emph{pairwise} distribution with $X^n$ and with $Y_i^n$ does not depend on  the hypothesis $\mathcal{H}$. In fact, under the null hypothesis
\begin{align}
&\mathcal{H}=0\colon \nonumber\\& (X^n,Y_1^n,Y_2^n,Z_1^n,Z_2^n)\sim \text{i.i.d.}\;P_{XY_1Y_2Z_1Z_2},
\end{align}
and under the alternative hypothesis,
\begin{align}
&\mathcal{H}=1\colon \nonumber\\& (X^n,Y_1^n,Y_2^n,Z_1^n,Z_2^n)\sim \text{i.i.d.}\; P_{XZ_1Z_2}P_{Y_1|Z_1}P_{Y_2|Z_2}.\label{h1c}
\end{align}
Here $P_{XY_1Y_2Z_1Z_2}$ is a given joint distribution over a finite product alphabet $\mathcal{X}\times \mathcal{Y}_1\times \mathcal{Y}_2\times \mathcal{Z}_1\times \mathcal{Z}_2$, and $P_{XZ_1Z_2}$, $P_{Y_1|Z_1}$ and $P_{Y_2|Z_2}$ denote its conditional marginals, i.e.,

\begin{align}
&P_{XZ_1Z_2}(x,z_1,z_2)\nonumber\\ &=\sum_{y_1\in\mathcal{Y}_1,y_2\in\mathcal{Y}_2}P_{XZ_1Z_2Y_1Y_2}(x,z_1,z_2,y_1,y_2),\nonumber\\[-2ex]&\qquad\qquad\qquad\qquad\;\; (x,z_1,z_2)\in\mathcal{X}\times \mathcal{Z}_1\times \mathcal{Z}_2,\nonumber\\[1ex]
&P_{Y_1|Z_1}(y_1|z_1) \nonumber\\&= \sum_{x\in\mathcal{X},y_2\in\mathcal{Y}_2,z_2\in\mathcal{Z}_2} P_{XY_1Y_2Z_2|Z_1}(x,y_1,y_2,z_2|z_1),\nonumber\\[-2ex]&\qquad\qquad\qquad\qquad\qquad (y_1,z_1)\in\mathcal{Y}_1\times \mathcal{Z}_1,\nonumber\\[1ex]
&P_{Y_2|Z_2}(y_2|z_2) \nonumber\\&=\sum_{x\in\mathcal{X},y_1\in \mathcal{Y}_1,z_1\in\mathcal{Z}_1}P_{XY_1Y_2Z_1|Z_2}(x,y_1,y_2,z_1|z_2),\nonumber\\[-2ex]&\qquad\qquad\qquad\qquad\qquad (y_2,z_2)\in  \mathcal{Y}_2\times \mathcal{Z}_2.\nonumber\\ \nonumber
\end{align}
The test here is ``against conditional independence" because $Z_i$ has the same joint distribution with the source $X$ under both hypotheses and because under $\mathcal{H}=1$, $Y_i$ is conditionally independent of $X$  given $Z_i$.

The transmitter communicates with the two receivers over 1 common and 2 individual noise-free bit pipes. Specifically, it
computes messages $(M_0,M_1,M_2)=\phi^{(n)}(X^n)$, using a possibly stochastic encoding function $\phi^{(n)}$ of the form $\phi^{(n)}:\mathcal{X}^n\to \{0,...,2^{nR_0}\}\times \{0,...,2^{nR_1}\}\times \{0,...,2^{nR_2}\}$, and sends message $M_0$ over the common pipe and messages $M_1$ and $M_2$ over the two individual pipes. For $i\in\{1,2\}$, Receiver~$i$ observes messages $M_0$ and $M_i$ and decides on the hypothesis $\mathcal{H}\in \{0,1\}$ by means of a decoding function 
$g_i^{(n)}\colon \mathcal{Y}_i^n\times \mathcal{Z}_i^n\times \{0,...,2^{nR_0}\}\times \{0,...,2^{nR_i}\}\to \{0,1\}$.
It produces $\hat{\mathcal{H}}_i=g_i^{(n)}(Y_i^n,Z_i^n,M_0,M_i)$.

\begin{definition}\label{defHB}
	For each $\epsilon \in (0,1)$, an exponents-rates tuple $(\theta_1,\theta_2,R_0,R_1,R_2)$ is called \emph{ $\epsilon$-achievable over the Gray-Wyner network with side information}  if there exists a sequence of encoding and decoding functions $\{(\phi^{(n)}, g_1^{(n)},$ $ g_2^{(n)})\}_{n=1}^\infty$ such that for $i\in\{1,2\}$ and all positive integers $n$, the corresponding sequences of type-I error probabilities 	
	\begin{equation}
	\label{eq:ai}	\alpha_{i,n}\stackrel{\Delta}{=} \Pr[\hat{\mathcal{H}}_i=1|\mathcal{H}=0],
	\end{equation}
	and type-II error probabilities
	\begin{equation}\label{eq:bi}
	\beta_{i,n}\stackrel{\Delta}{=} \Pr[\hat{\mathcal{H}}_i=0|\mathcal{H}=1],
	\end{equation}
	satisfy $$\alpha_{i,n}\leq \epsilon,$$ and $$-\varlimsup_{n\to \infty}\frac{1}{n}\log\beta_{i,n}\geq \theta_i.$$
\end{definition}
\begin{definition} Given nonnegative rates $(R_0,R_1,R_2)$, define the exponents region $\mathcal{E}_{\GW}^{\SI}(R_0,R_1,R_2)$ as the closure of all non-negative exponent pairs $(\theta_1,\theta_2)$ for which $(\theta_1,\theta_2,R_0,R_1,R_2)$ is $\epsilon$-achievable over the Gray-Wyner network with side information for every $\epsilon\in (0,1)$.
\end{definition}

\begin{remark}\label{remregion} The exponents region $\mathcal{E}_{\GW}^{\SI}(R_0,R_1,R_2)$ only depends on the marginal distributions $P_{XZ_1Z_2}$, $P_{XY_1|Z_1}$ and $P_{XY_2|Z_2}$ under both hypotheses. 
\end{remark}

\subsection{Coding and Testing Scheme}\label{sec:scheme2}


We propose to split the block of $n$ transmissions into  $B$  subblocks of $k$ consecutive transmissions each such that $n=kB$. So, for each $b\in\{1,\ldots, B\}$, let 
\begin{IEEEeqnarray}{rCl}
X_b^{k}& := &(X_{(b-1)k +1},\ldots, X_{bk}),\label{xblock}\\
Y_{i,b}^{k}& := &(Y_{i,(b-1)k +1},\ldots, Y_{i,bk}), \;\;\; i\in\{1,2\},\label{yblock}\\
Z_{i,b}^{k}& := &(Z_{i,(b-1)k +1},\ldots, Z_{i,bk}), \;\;\; i\in\{1,2\}.\label{zblock}
\end{IEEEeqnarray} 
For each of the subblocks, we propose to apply an independent instance of the  coordination code for the Gray-Wyner network with side-information  in \cite{Shayevitz}, where  the receivers  only account for side-informations $Z_1^n$ and $Z_2^n$ but not for $Y_1^n$ and $Y_2^n$.  More specifically, choose a small real number 
$\mu>0$, as well as
  auxiliary alphabets $\mathcal{U}_0$, $\mathcal{U}_1$, and $\mathcal{U}_2$, and
	a conditional joint probability distribution $P_{U_0 U_1 U_2|X}$ over $\mathcal{U}_0\times \mathcal{U}_1\times \mathcal{U}_2$ so that 

	\begin{IEEEeqnarray}{rCl}
	R_0 + R_1 & \geq & I(U_0,U_1;X|Z_1)+\mu, \\
	R_0 +R_2 & \geq &I(U_0,U_2;X|Z_2)+\mu,\\
	R_0+R_1+R_2 & \geq & \max_{i\in\{1,2\}} I(U_0;X|Z_i) \nonumber\\&&\;\;+I(U_1;X|U_0,Z_1)\nonumber\\&&\;\;+I(U_2;X|U_0,Z_2)+\mu.
	\end{IEEEeqnarray}
Construct for each block a coordination code as described in \cite[Section~V-B1)]{Shayevitz} for suitably chosen auxiliary rates $R_{0,0}, R_{0,1}, R_{0,2}, R_{1,0}, R_{1,1}, R_{2,0}, R_{2,2}, R_0', R_1', R_2'>0$ satisfying $R_{0}'> \max\{R_{1,0},R_{2,0}\}$ and Constraints~(50) in \cite[Appendix~B]{Shayevitz}.


\textbf{Codebook Generation}: Let $P_{U_0}$, $P_{U_1|U_0}$ and $P_{U_2|U_0}$ be the marginal and conditional marginal pmfs of $P_{X} \cdot P_{U_0U_1U_2|X}$. 

For each block $b\in\{1,\ldots, B\}$,  generate three codebooks $\mathsf{C}_{0,b}, \mathsf{C}_{1,b}(.), \mathsf{C}_{2,b}(.)$ independently of each other in the following way.
Codebook $\mathsf{C}_{0,b}$ consists of $ 2^{kR_{0,0}}$ superbins, each containing $ 2^{k{R}_{0}'} $  length-$k$ codewords whose entries are  randomly and independently generated according to  the law $P_{U_0}$. 

We make two partitions of the codewords in each superbin. In the first partition, the codewords of each superbin are assigned to 
$ 2^{k{R}_{1,0}}$ subbins, each containing $ 2^{k({R}_0'-R_{1,0})}$ codewords; in the second partition they are assigned to
$ 2^{k{R}_{2,0}}$ subbins, each containing $ 2^{k({R}_0'-R_{2,0})}$ codewords. There are thus two different ways to refer to a specific codeword in $\mathsf{C}_{0,b}$.
When we consider the first partition, we denote the  
codewords in the $m_{1,0,b}\in \{1,\ldots,2^{k{R}_{1,0}}\}$-\textit{th}
subbin of  superbin $m_{0,0,b}\in\{1,\ldots,2^{kR_{0,0}}\}$ by 
\[\{u_{0,b}^k(1; m_{0,0,b}, m_{1,0,b},\ell_{1,0,b})\}_{\ell_{1,0,b}=1}^{ 2^{k({R}_0'-R_{1,0})}};\]
when we consider the second  partition, we denote the  codewords in the   $m_{2,0,b}\in \{1,\ldots,2^{k{R}_{2,0}}\}$-\textit{th} subbin of  superbin 
$m_{0,0,b}\in\{1,\ldots,2^{kR_{0,0}}\}$ by 
\[\{u_{0,b}^k(2;m_{0,0,b}, m_{2,0,b},\ell_{2,0,b})\}_{\ell_{2,0,b}=1}^{ 2^{k({R}_0'-R_{2,0})} }.\]
Thus, here the first index indicates whether the last two indices refer to the first or the second partition of the superbins.

For $i\in\{1,2\}$, Codebook $\mathsf{C}_{i,b}(.)$ consists of  $ 2^{kR_{0,i}}$ superbins each containing $ 2^{kR_{i,i}}$ subbins with $ 2^{kR_{i}'}$ codewords of length $k$, where all entries of all codewords are randomly and independently drawn according to $P_{U_i}$.
For $m_{i,i,b}\in\{1,\ldots,2^{kR_{i,i}}\}$, we denote the codewords in the $m_{i,i,b}$-\textit{th} subbin of superbin $m_{0,i,b} \in 2^{kR_{0,i}}$ by \[\{u_{i,b}^k(m_{0,i,b}, m_{i,i,b}, \ell_{i,b})\}_{\ell_{i,b}=1}^{ 2^{kR_i'} }.\]

All codebooks are revealed to the sender, and codebooks $\{\mathsf{C}_{0,b},\mathsf{C}_{i,b}(.)\}$  are revealed to Receiver~$i\in\{1,2\}$.

\textbf{Transmitter:}
The transmitter first decomposes the observed source sequence $X^n=x^n$ into $B$ blocks, each consisting of $k$ consecutive symbols, $x_1^{k},\ldots, x_{B}^k$.  For each block $b\in\{1,\ldots, B\}$, it then forms a list of all the tuples of indices $(m_{0,0,b}, m_{1,0,b},  \ell_{1,0,b}, m_{0,1,b}, m_{1,1,b}, \ell_{1,b}, m_{0,2,b},$ $  m_{2,2,b},\ell_{2,b})$ so that the 
 triplet of codewords $u_{0,b}^{k}(1;m_{0,0,b}, m_{1,0,b}, \ell_{1,0})\in\mathsf{C}_{0,b}$, $u_{1,b}^k(m_{0,1,b}, m_{1,1,b},$ $\ell_{1,b})\in\mathsf{C}_{1,b}(.)$, $u_{2,b}^k(m_{0,2,b}, m_{2,2,b},\ell_{2,b})\in\mathsf{C}_{2,b}(.)$ satisfies
 
\begin{align}
&(x_b^k,u_{0,b}^k(1;m_{0,0,b}, m_{1,0,b}, \ell_{1,0,b}),\nonumber\\& u_{i,b}^k(m_{0,i,b}, m_{i,i,b},\ell_{i,b}))\in\mathcal{T}_{\mu/2}^k(P_{XU_0U_i}), \quad \in\{1,2\}.\label{eq:typcond}
\end{align}

If for some block $b$ this list is empty, the transmitter sends  the messages $m_{0}=0$, $m_{1}=0$ and $m_{2}=0$ over the bit pipes. Otherwise, it 
 chooses for each block $b$ the  tuple $(m_{0,0,b}^\star, m_{1,0,b}^\star,  \ell_{1,0,b}^\star,$ $m_{0,1,b}^\star, m_{1,1,b}^\star,\ell_{1,b}^\star, m_{0,2,b}^\star, m_{2,2,b}^\star,\ell_{2,b}^\star)$ uniformly at random over the generated list, and sends the following messages over the bit pipes
\begin{align}m_0&=(m_{0,0,1}^\star,\ldots,m_{0,0,B}^\star, m_{0,1,1}^\star, \ldots, m_{0,1,B}^\star,\nonumber\\&\qquad m_{0,2,1}^\star, \ldots, m_{0,2,B}^\star),\\
m_1 &= (m_{1,0,1}^\star,\ldots,m_{1,0,B}^\star,m_{1,1,1}^\star,\ldots,m_{1,1,B}^\star),\\
m_2 &= (m_{2,0,1}^\star,\ldots,m_{2,0,B}^\star,m_{2,2,1}^\star,\ldots,m_{2,2,B}^\star).
\label{message0}\end{align}

\textbf{Receiver~$i$:}
Assume that Receiver~$i$ observes messages $M_0=m_0$, $M_i=m_i$ and source sequences $Y_i^n=y_i^n$ and $Z_i^n=z_i^n$. If $m_0=m_i=0$, Receiver~$i$ declares $\hat{\mathcal{H}}_i=1$. Otherwise, it decomposes its observations into $B$ blocks

\begin{align}\big\{ \big(m_{0,b},m_{i,b}, y_{i,b}^k, z_{i,b}^k\big)\big\}_{b=1}^B.\end{align}
 It parses the common message $m_{0,b}$ as $(m_{0,0,b},m_{0,1,b},m_{0,2,b})$ and its private message $m_{i,b}$ as $m_{i,b}=(m_{i,0,b},m_{i,i,b})$. Then, it seeks a codeword $u_{0,b}^k(i;m_{0,0,b},m_{i,0,b},\ell_{i,0,b})$ in codebook $\mathsf{C}_{0,b}$ and a codeword $u_{i,b}^k(m_{0,i,b},m_{i,i,b},\ell_{i,b})$ in codebook $\mathsf{C}_{i,b}(.)$ such that 
 \begin{align}
 &\big( u_{0,b}^k(i;m_{0,0,b},m_{i,0,b},\ell_{i,0,b}), u_{i,b}^k(m_{0,i,b},m_{i,i,b},\ell_{i,b}),\nonumber\\&\qquad\qquad\qquad\qquad z_{i,b}^k \big)\in \mathcal{T}_{\mu}^k(P_{U_0U_iZ_i}).
 \end{align}
If exactly one such pair of codewords exists, Receiver $i$  produces the coordination sequence $\hat{u}_{i,b}^k=u_{i,b}^k(m_{0,i,b},m_{i,i,b},\ell_{i,b})$. Otherwise, it randomly chooses a triplet $(m^*_{0,i,b},m^*_{i,i,b},\ell^*_{i,b})$ and produces the coordination sequence $\hat{u}_{i,b}^k=u_{i,b}^k(m^*_{0,i,b},m^*_{i,i,b},\ell^*_{i,b})$. Finally, it applies a Neyman-Pearson test to decide on hypothesis $\mathcal{H}$ based on the i.i.d. sequence of tuples 
\begin{equation}
\big\{ \big(\hat{u}_{i,b}^k, y_{i,b}^k, z_{i,b}^k\big)\big\}_{b=1}^B,\label{coorseq}
\end{equation}
in a way that the type-I error probability  does not exceed $\epsilon$.

\subsection{Result on Exponents Region}\label{GWSIachsec}

{
The scheme described in the previous section gives the following achievable exponents region.

Let  $\mathcal{E}_{\text{\GW}}^{\SI,\text{in}}(R_0,R_1,R_2)$  be given by the following:

	\begin{align}
		&\mathcal{E}_{\GW}^{\SI,\text{in}}(R_0,R_1,R_2) := \nonumber\\& \!\!\!\!\!\!\!\!\!\!\!\!\!\!\!\!\!\!\bigcup_{\substack{(U_0,U_1,U_2)\colon \\[1ex]
			(U_0,U_1,U_2)\to X\\\;\;\;\to (Y_1,Y_2,Z_1,Z_2)\\[1ex]
		R_0+R_1+R_2\geq  \max_{i\in\{1,2\}} \\\hspace{3cm} I(U_0;X|Z_i) \nonumber \\
		\hspace{3.5cm}+I(U_1;X|U_0,Z_1)\\\hspace{3.5cm}+I(U_2;X|U_0,Z_2)\\[1ex]\qquad \; R_0+R_1 \geq I(U_1,U_0;X|Z_1)\\[1ex]\qquad \; R_0+R_2 \geq I(U_0,U_2;X|Z_2)	
	}} \!\!\!\!\!\!\!\! \!\!\!\!\!\!\!\! \!\!\!\!\!\!\!\! \!\!\!\!\!\!\!\! \left\{ \begin{array}{ll}  (\theta_1,\theta_2) \colon  \theta_1 \geq 0, \theta_2\geq 0, \\\;\;\begin{array}{l} \qquad    \;\;\;\theta_1 \leq I(U_1;Y_1|Z_1) \\ \qquad   \;\;\;	\theta_2 \leq I(U_2;Y_2|Z_2)
	\end{array} \end{array} \right\}.\label{thmina}
	\end{align}
\vspace{3mm}

Notice that, to evaluate $\mathcal{E}_{\GW}^{\SI,\text{in}}(R_0,R_1,R_2)$  it suffices to consider auxiliary random variables $U_0, U_1, U_2$ over alphabets $\mathcal{U}_0$, $\mathcal{U}_1$, and $\mathcal{U}_2$ whose sizes satisfy the following three conditions: $|\mathcal{U}_0| \leq |\mathcal{X}| +3$, $|\mathcal{U}_1|\leq |\mathcal{X}| \cdot |\mathcal{U}_0|+1$, and $|\mathcal{U}_2|\leq |\mathcal{X}| \cdot |\mathcal{U}_0|+1$.

 \begin{theorem}\label{mainthm} The set $\mathcal{E}_{\GW}^{\SI,\text{in}}(R_0,R_1,R_2)$  is achievable, i.e., 
 \begin{equation}
 \mathcal{E}_{\GW}^{\SI,\text{in}}(R_0,R_1,R_2) \subseteq \mathcal{E}_{\GW}^{\SI}(R_0,R_1,R_2).
 \end{equation}
 \end{theorem}
 	\begin{IEEEproof}
	See Appendix~\ref{app:1}.
	\end{IEEEproof}
}

 The two next-following results show that the exponents region $\mathcal{E}_{\GW}^{\SI,\text{in}}$ coincides with the optimal exponents region $\mathcal{E}_{\GW}^{\SI}$ in some special cases. 
%
%
%
 
 Let
  \begin{align}
 		&\mathcal{E}_{\GW}(R_0,R_1,R_2) :=\nonumber\\& \bigcup_{\substack{(U_0,U_1,U_2)\colon \\[1ex]
 				(U_0,U_1,U_2)\to X\to (Y_1,Y_2)\\[1ex]
 				R_0 \geq I(U_0;X)\\[1ex] R_1 \geq I(U_1;X|U_0)\\[1ex] \; R_2 \geq I(U_2;X|U_0)}} \left\{ \begin{array}{ll}  (\theta_1,\theta_2) \colon  \theta_1 \geq 0, \theta_2\geq 0, \\ \begin{array}{l} \qquad  \; \;\;\;	\theta_1 \leq I(U_1;Y_1) \\ \qquad\; \;\;\;	\theta_2 \leq I(U_2;Y_2) 
 		\end{array}  \end{array} \right\}.\label{thm1a}
 		\end{align}
%
	\vspace{3mm}	
 
 \begin{theorem}\label{thm1} When there is no side-information, i.e., $Z_1$ and $Z_2$ are constants, 
 then 
 \begin{align} 		
 	 		\mathcal{E}_{\GW}^{\textnormal{SI}}(R_0,R_1,R_2) &= \mathcal{E}_{\GW}^{\textnormal{SI,in}}(R_0,R_1,R_2)\nonumber\\&\qquad\qquad= \mathcal{E}_{\GW}(R_0,R_1,R_2).
 		\end{align}
 \end{theorem} 
 \begin{IEEEproof} Achievability follows by specializing Theorem \ref{mainthm}  to $Z_1$ and $Z_2$ constant.  The converse can be obtained from the converse in \cite{Michele2} where one has to include $U_0$ into $U_1$. \end{IEEEproof}
 
 In the above Theorem~\ref{thm1} it suffices to consider auxiliary random variables $U_0$, $U_1$, and $U_2$ over alphabets $\mathcal{U}_0$, $\mathcal{U}_1$, and $\mathcal{U}_2$ whose sizes satisfy:  
 \begin{IEEEeqnarray}{rCl}
 |\mathcal{U}_0|& \leq& |\mathcal{X}| + 2, \\
 |\mathcal{U}_j| &\leq& |\mathcal{X}|\cdot|\mathcal{U}_0|+1, \qquad j\in\{1,2\}. 
 \end{IEEEeqnarray}
  This follows by simple applications of Caratheodory's theorem.

 \begin{theorem}\label{thm3}Let $Z_2$ be a constant and $Z_1$ {less noisy} than $Y_2$, i.e.,  let for all auxiliary random variables~$U$ satisfying the Markov chain $U\to X\to (Y_1,Y_2,Z_1)$ the following inequality hold:
 	\begin{align}\label{eq:ineq}
 	I(U;Z_1)\geq I(U;Y_2).
 	\end{align}
	Then:
	\begin{align}
 	&\mathcal{E}_{\GW}^{\textnormal{SI}}(R_0,R_1=0,R_2=0) =\nonumber\\&\qquad\qquad\qquad \mathcal{E}_{\GW}^{\textnormal{SI,in}}(R_0,R_1=0,R_2=0).
	\end{align}
 \end{theorem}
 \begin{IEEEproof} Achievability follows by Theorem \ref{mainthm}. The converse is proved in Appendix \ref{thm3p}. 
 \end{IEEEproof}
 
 
 \subsection{An Example}\label{sec:example}

 Theorem \ref{thm3} was stated for discrete memoryless sources. {It can be shown that it remains valid also when sources are memoryless and jointly Gaussian \cite[Chap. 3]{ElGamal}.}
 
 Consider the following scenario. Under both hypotheses, $X\sim \mathcal{N}(0,1)$ and 
 $Z_1= X+ N_z$,
 where $N_z\sim\mathcal{N}(0,\sigma_z^2)$ is independent of $X$.  Moreover, under hypothesis
 \begin{align}
 \mathcal{H}=0 \colon \qquad Y_1 &= X + Z_1+  N_1,\label{ex1}\\
 Y_2 &= Z_1 + N_2,\label{ex2}
 \end{align}
 where $N_1\sim \mathcal{N}(0,\sigma_1^2)$ and $N_2\sim \mathcal{N}(0,\sigma_2^2)$ are independent of each other and of $(X,Z_1)$, and under hypothesis
 \begin{align}
 \mathcal{H}=1 \colon \qquad  \qquad Y_1 &= X' +\frac{2+\sigma_z^2}{1+\sigma_z^2} \cdot Z_1 +N_1,\label{ex3}\\
 Y_2&= Z'_1 +N_2,
 \end{align}
 where $X'\sim \mathcal{N}(0,\frac{\sigma_z^2}{1+\sigma_z^2})$ and $Z_1'\sim \mathcal{N}(0,1+\sigma_z^2)$ are independent of each other and of the tuple $(X, Z_1, N_1, N_2)$.

 The described scenario satisfies the less noisy condition in \eqref{eq:ineq}. By Theorem~\ref{thm3}, when restricting to $R_1=R_2=0$,  for this example, the region $\mathcal{E}_{\GW}^{\SI}$  equals $\mathcal{E}_{\GW}^{\SI,\textnormal{in}}$. 
 As is proved in Appendix~\ref{app:example},  
 the exponents region $\mathcal{E}_{\GW}^{\SI}(R_0,R_1=0,R_2=0)$  evaluates to the set of all nonnegative exponent pairs $(\theta_1,\theta_2)$ that satisfy 
 \begin{subequations}\label{eq:subeqs}
 	\begin{IEEEeqnarray}{rCl}
 		\theta_1 &\leq& \frac{1}{2} \log \bigg( \frac{{\sigma^2_z} +\sigma_1^2(1+\sigma^2_z)}{2^{2 \tilde{\alpha}} {\sigma^2_z} +\sigma_1^2(1+\sigma^2_z)}\bigg), \\
 		\theta_2 & \leq&\frac{1}{2} \log \bigg(\frac{ 1 +\sigma^2_z+ \sigma_2^2 }{2^{-2 (\tilde{\alpha}+R_0)} (1+\sigma^2_z) +\sigma_2^2}\bigg),	\end{IEEEeqnarray}
 \end{subequations} 
 for some $\tilde{\alpha} \in [ -R_0, 0]$.
 
  \begin{figure}[t]
 	\centering
 	\includegraphics[scale=0.4,angle=0]{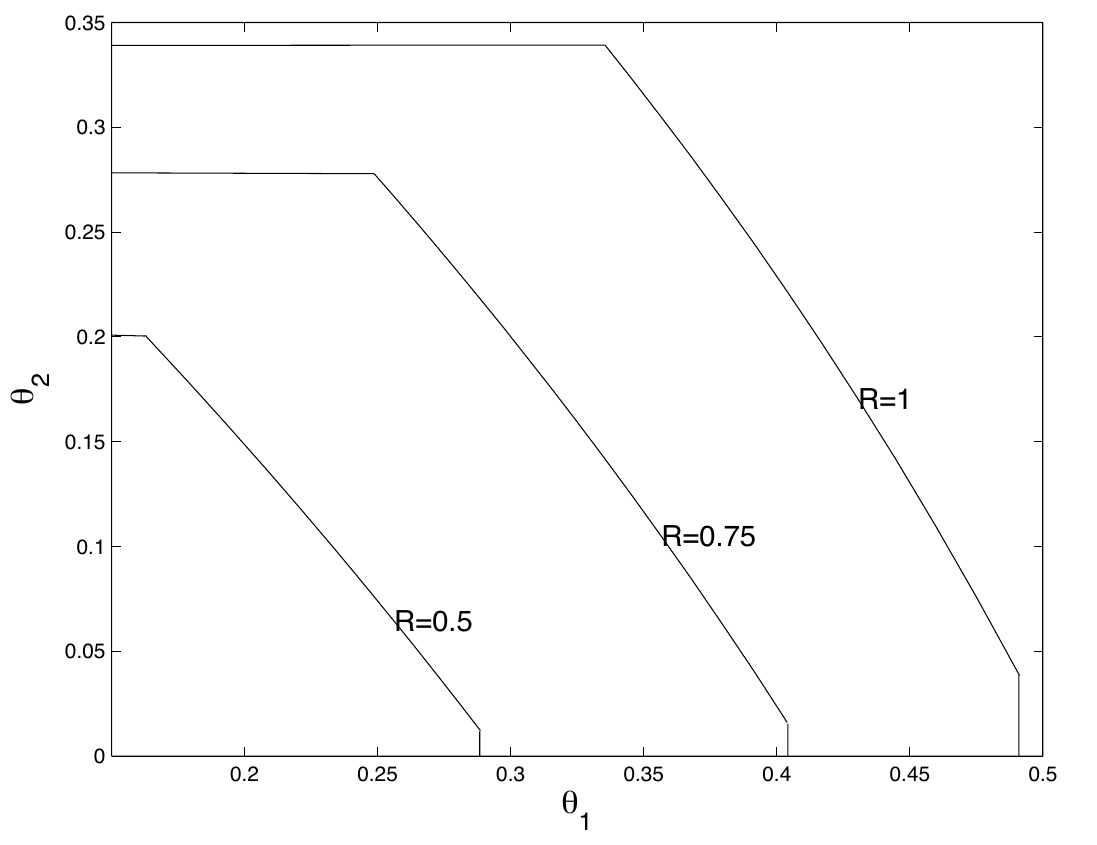}
 	\caption{Exponents region for $\sigma_z^2=0.7$, $\sigma_1^2=0.2$ and $\sigma_2^2=0.3$.}
 	\vspace{0cm}
 	\label{Figure 5}
 \end{figure}
 The boundary of the exponents  region $\mathcal{E}_{\GW}^{\SI}(R_0,R_1=0,R_2=0)$  is illustrated in Fig.~\ref{Figure 5} for different values of the rate $R_0$. Generally, on this boundary $\theta_1>\theta_2$, because Receiver~1 has the additional side-information $Z_1$. One  observes a trade-off between the two exponents $\theta_1$ and $\theta_2$, which is captured by the parameter $\tilde{\alpha}$ in \eqref{eq:subeqs}. In other words, having a larger exponent $\theta_1$ comes at the expense of a smaller exponent $\theta_2$, and vice versa.

\section{Hypothesis Testing over Noisy Channels}\label{secnoisy}
\begin{figure*}[t]
	\centering
	\includegraphics[scale=0.27,angle=0]{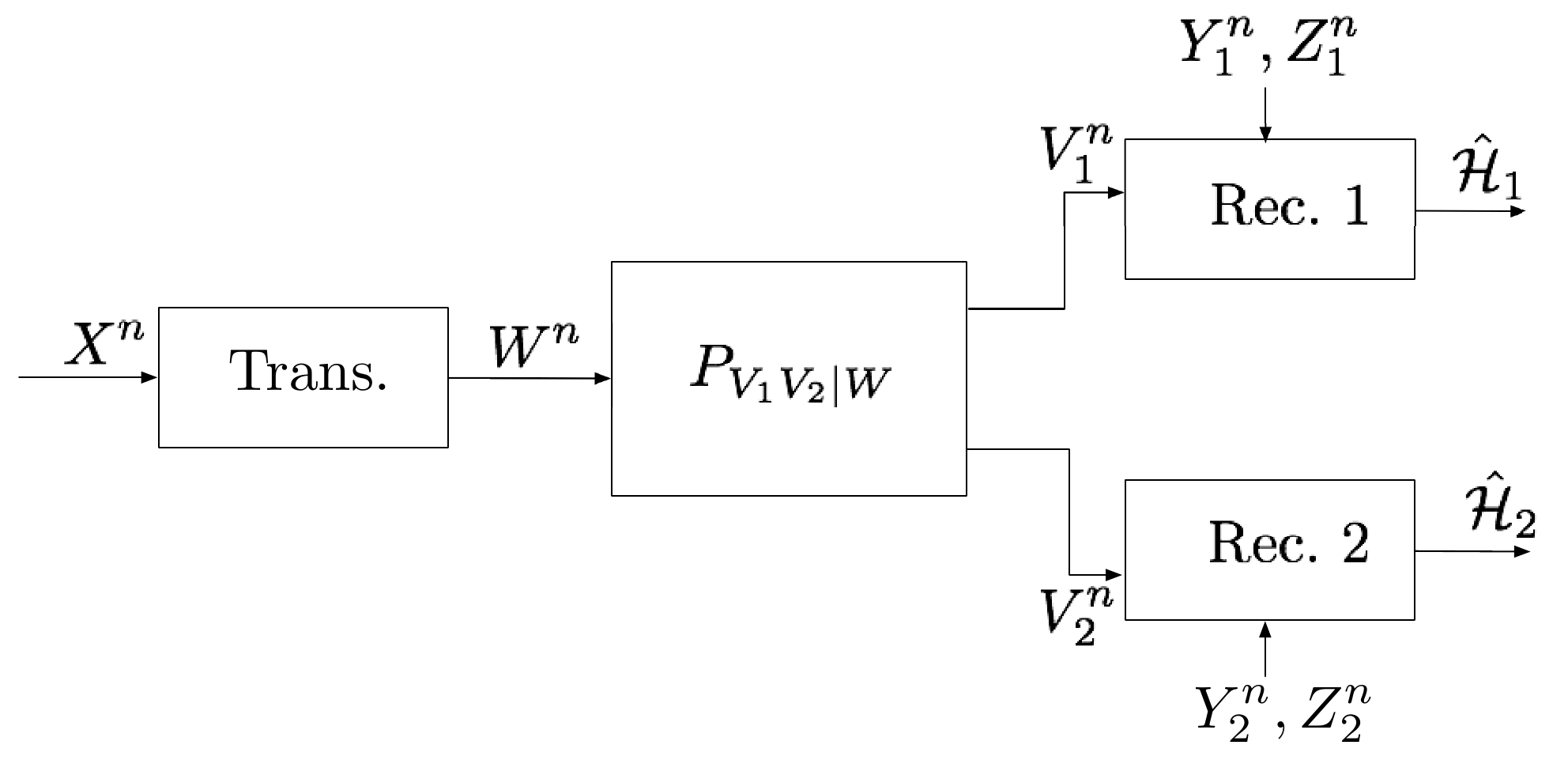}
	\caption{Hypothesis testing over a  BC.}
	\label{Figure 6}
\end{figure*}

This section considers hypothesis testing over a discrete memoryless BC $(\mathcal{W}, \mathcal{V}_1,\mathcal{V}_2, P_{V_1V_2|W})$, where $\mathcal{W}$ denotes the finite channel input alphabet, $\mathcal{V}_1$ and $\mathcal{V}_2$ the finite channel output alphabets at Receivers 1 and 2, and $P_{V_1V_2|W}$ the BC transition pmf. 
The setup is illustrated in Fig. \ref{Figure 6}. The transmitter observes a sequence $X^n$ and  produces its channel inputs $W^n:=(W_1,\ldots, W_n)$ as $W^n=\Phi^{(n)}(X^n)$ by means of a possibly stochastic encoding function $\Phi^{(n)} \colon \mathcal{X}^n \to \mathcal{W}^n$. Receivers~1 and 2 observe the corresponding channel outputs $V_1^n:=(V_{1,1},\ldots, V_{1,n})$ and $V_{2}^n:=(V_{2,1},\ldots, V_{2,n})$, as well as the source sequences $(Y_1^n, Z_1^n)$ and $(Y_2^n,Z_2^n)$ defined in the previous section. For $i\in\{1,2\}$, Receiver~$i$ decides on the hypothesis $\mathcal{H}\in \{0,1\}$ by means of a decoding function 
$g_i^{(n)}\colon \mathcal{Y}_i^n\times \mathcal{Z}_i^n\times \mathcal{V}_i^n\to \{0,1\}$.
It produces $\hat{\mathcal{H}}_i=g_i^{(n)}(Y_i^n,Z_i^n,V_i^n)$.

As in the previous section, assume that under hypothesis
\begin{align}
\mathcal{H}=0\colon (X^n,Y_1^n,Y_2^n,Z_1^n,Z_2^n)\sim \text{i.i.d.}\;P_{XY_1Y_2Z_1Z_2},
\end{align}
and under hypothesis
\begin{align}
\mathcal{H}=1\colon & (X^n,Y_1^n,Y_2^n,Z_1^n,Z_2^n)\sim \text{i.i.d.}\;\nonumber\\&\qquad  P_{XZ_1Z_2}P_{Y_1|Z_1}P_{Y_2|Z_2}.
\end{align}

\begin{definition}\label{defBC}
	For each $\epsilon \in (0,1)$, an exponent pair $(\theta_1,\theta_2)$ is called \emph{ $\epsilon$-achievable over a  BC with side information}  if there exists a sequence of encoding and decoding functions $\{(\Phi^{(n)}, g_1^{(n)},$ $ g_2^{(n)})\}_{n=1}^\infty$ such that for $i\in\{1,2\}$ and all positive integers $n$, the corresponding sequences of type-I and type-II error probabilities 	
	satisfy  $$\alpha_{i,n}\leq \epsilon,$$ and $$-\varlimsup_{n\to \infty}\frac{1}{n}\log\beta_{i,n}\geq \theta_i,$$
	where $\alpha_{i,n}$ and $\beta_{i,n}$ are defined in \eqref{eq:ai} and \eqref{eq:bi}.
\end{definition}
\begin{definition} Define the exponents region $\mathcal{E}_{BC}^{\SI}$ as the closure of all non-negative exponent pairs $(\theta_1,\theta_2)$ for which $(\theta_1,\theta_2)$ is $\epsilon$-achievable over the BC with side information for every $\epsilon\in (0,1)$.
\end{definition}

\subsection{Coding and Testing Scheme}\label{sec:scheme4}


Fix $\mu>0$, sufficiently large positive integers $k$ and $B$, and a joint conditional distribution $P_{U_0U_1U_2|X}$ over finite auxiliary alphabets $\mathcal{U}_0$, $\mathcal{U}_1$ and $\mathcal{U}_2$.  Consider also  nonnegative rates $R_0,R_1,R_2$ that satisfy 
\begin{align}
R_0+R_1 &\leq I(U_1,U_0;V_1,Z_1),\label{eq:rc1}\\
R_0+R_2 &\leq I(U_2,U_0;V_2,Z_2),\label{eq:rc5}\\
R_1 &\leq I(U_1;V_1,Z_1|U_0),\label{eq:rc6}\\
R_2 &\leq I(U_2;V_2,Z_2|U_0),\label{eq:rc8}\\
R_0 &>I(U_0;X),\label{eq:rc4}\\
R_1 &>I(U_1;X|U_0),\label{eq:rc9}\\
R_2 &>I(U_2;X|U_0),\label{eq:rc7}\\
R_1+R_2 &>I(U_1,U_2;X|U_0)+I(U_1;U_2|U_0).\label{rc8}
\end{align}
Finally, fix a function $f\colon \mathcal{U}_0 \times \mathcal{U}_1\times \mathcal{U}_2 \times \mathcal{X} \to \mathcal{W}$.

\textbf{Code Construction}: For each block $b\in \{1,...,B\}$, randomly generate a codebook $\mathsf{C}_{0,b}=\{U_{0,b}^k(m_{0,b}):m_{0,b}\in\{1,...,2^{kR_0}\} \}$ by  drawing each entry of the $n$-length codeword $U_{0,b}^k(m_{0,b})$  i.i.d. according to the  pmf $P_{U_0}$. Moreover, for each index $m_{0,b}$ and $i\in\{1,2\}$, randomly generate a codebook $\mathsf{C}_{i,b}(m_{0,b}) := \{U_{i,b}^k(m_{i,b}|m_{0,b})\colon m_{i,b}\in\{1,...,2^{kR_i}\}\}$ by drawing each entry of the $k$-length codeword $U_{i,b}^k(m_{i,b}|m_{0,b})$ i.i.d. according to the conditional  pmf $P_{U_i|U_0}(.|U_{0,b,j}(m_{0,b}))$, where $U_{0,b,j}(m_{0,b})$ denotes the $j$-\textit{th} symbol of $U_{0,b}^k(m_{0,b})$. 
Reveal the realizations $\{\mathcal{C}_{0,b}\}$, $\{\mathcal{C}_{1,b}(\cdot)\}$ and $\{\mathcal{C}_{2,b}(\cdot)\}$ of the randomly generated codebooks  to all terminals.

\textbf{Transmitter}: It observes a source sequence $x^n$ and splits it into $B$ subblocks $x^n=(x_1^k,...,x_B^k)$ as in \eqref{xblock}. For each block $b$, it looks for a triple of indices $(m_{0,b},m_{1,b},m_{2,b})\in\{1,\ldots, 2^{kR_0}\} \times \{1, \ldots, 2^{kR_1}\}\times \{1, \ldots, 2^{kR_2}\}$ such that 

\begin{align}&(x_b^k,u_{0,b}^k(m_{0,b}),u_{1,b}^k(m_{1,b}|m_{0,b}),u_{2,b}^k(m_{2,b}|m_{0,b}))\nonumber\\&\qquad\qquad\qquad\qquad\qquad\in \mathcal{T}_{\mu/2}^k(P_{XU_0U_1U_2}),\end{align}

where $u_{0,b}^k(m_{0,b})$, $u_{1,b}^k(m_{1,b}|m_{0,b})$ and $u_{2,b}^k(m_{2,b}|m_{0,b})$ are codewords from the chosen codebooks $\mathcal{C}_{0,b}$, $\{\mathcal{C}_{1,b}(\cdot)\}$ and $\{\mathcal{C}_{2,b}(\cdot)\}$.
If the typicality test is successful, the transmitter picks one of the triples satisfying the test at random. Otherwise, it picks a triple $(m_{0,b}, m_{1,b},m_{2,b})$ uniformly at random over $\{1,\ldots, 2^{kR_0}\} \times \{1, \ldots, 2^{kR_1}\}\times \{1, \ldots, 2^{kR_2}\}$.
It finally sends  the $k$ inputs
\begin{align}w_{(b-1)k+j} &=f(u_{0,b,j}(m_{0,b}),u_{1,b,j}(m_{1,b}|m_{0,b}),\nonumber\\&\qquad u_{2,b,j}(m_{2,b}|m_{0,b}),x_{(b-1)k+j}),\nonumber\\&\qquad\qquad\qquad \qquad j\in\{1,\ldots, k\},\end{align}
over the channel.

\textbf{Receiver $i\in\{1,2\}$}: Assume that it observes the sequence of channel outputs $v_{i,b}^n$ and the source sequences $y_{i,b}^n$ and $z_{i,b}^n$. It looks for a pair of indices $(\hat{m}_{0,b},\hat{m}_{i,b})$ such that 
\begin{align}
(u_{i,b}^k(\hat{m}_{i,b}|\hat{m}_{0,b}),v_{i,b}^k, z_{i,b}^k) \in \mathcal{T}_{\mu}^k(P_{U_iV_iZ_i}),
\end{align}
and picks one of these pairs at random. If no such pair can be found, pick $(\hat{m}_{0,b},\hat{m}_{i,b})$ uniformly over $\{1,\ldots,2^{kR_0}\}\times \{1,\ldots,2^{kR_1}\}$. For the chosen $(\hat{m}_{0,b},\hat{m}_{i,b})$, set \begin{equation}
\hat{u}_{i,b}^k := u_{i,b}^k(\hat{m}_{0,b},\hat{m}_{i,b}).
\end{equation}Receiver $i$ then decomposes its observations $(y_{i,b}^k,z_{i,b}^k)$ as in \eqref{yblock} and \eqref{zblock} and performs a Neyman-Pearson test on the $B$ i.i.d. blocks,
$$\big\{ \big(\hat{u}_{i,b}^k, v_{i,b}^k, y_{i,b}^k, z_{i,b}^k\}\big)\big\}_{b=1}^B,$$
in a way that the type-I error probability  does not exceed $\epsilon$.

\subsection{Exponents Region}

Let $\mathcal{E}_{BC}^{\text{hyb}}$ be given by the following:

\begin{align}
\mathcal{E}_{BC}^{\text{hyb}} &= \bigcup_{\substack{(U_0,U_1,U_2)
}}\left\{ \begin{array}{ll}  (\theta_1,\theta_2) \colon  \theta_1 \geq 0, \;\;\theta_2\geq 0,  \\\;\;\begin{array}{l} \qquad   \;\;\;\theta_1 \leq I(U_1;Y_1|Z_1) \\ \qquad\;\;\;	\theta_2 \leq I(U_2;Y_2|Z_2)
\end{array}   \end{array} \right\} ,\label{thmhybrid}\nonumber\\ \nonumber
\end{align}
where the union is taken over all pmfs $P_{U_0U_1U_2W|X}$ that satisfy the following Markov chains 
\begin{align}
&(U_0,U_1,U_2)\to X\to (Y_1,Y_2,Z_1,Z_2),\\
&(Y_1,Y_2,Z_1,Z_2)\to (U_0,U_1,U_2,X)\nonumber\\&\qquad\qquad\qquad \qquad\qquad\to W\to (V_1,V_2),
\end{align}
and the mutual information constraints
\begin{subequations}
	\label{eq:conshyb}
\begin{align}
&I(U_1,U_0;X|Z_1) \leq I(U_1,U_0;V_1|Z_1),\\ &I(U_2,U_0;X|Z_2)\leq I(U_2,U_0;V_2|Z_2),\\ & I(U_1;X|Z_1,U_0)\leq I(U_1;V_1|Z_1,U_0),\\ &I(U_2;X|Z_2,U_0)\leq I(U_2;V_2|Z_2,U_0),\\
&I(U_0,U_1;X|Z_1)+I(U_2;X|Z_2,U_0)+I(U_1;U_2|U_0) \nonumber\\&\qquad \leq I(U_0,U_1;V_1|Z_1)+I(U_2;V_2|Z_2,U_0),\\
&I(U_0,U_2;X|Z_2)+I(U_1;X|Z_1,U_0)+I(U_1;U_2|U_0) \nonumber\\&\qquad\leq I(U_1;V_1|Z_1,U_0)+I(U_0,U_2;V_2|Z_2),\\
&I(U_1;X|Z_1,U_0)+I(U_2;X|Z_2,U_0)+I(U_1;U_2|U_0) \nonumber\\&\qquad\leq I(U_1;V_1|Z_1,U_0)+I(U_2;V_2|Z_2,U_0),\\
&I(U_1,U_0;X|Z_1)+I(U_2,U_0;X|Z_2)+I(U_1;U_2|U_0) \nonumber\\&\qquad\leq I(U_1,U_0;V_1|Z_1)+I(U_2,U_0;V_2|Z_2),
\end{align}
\end{subequations}

for some function $f:\mathcal{U}_0\times \mathcal{U}_1\times \mathcal{U}_2 \times \mathcal{X}\to \mathcal{W}$ where $W=f(U_0,U_1,U_2,X)$.

\begin{theorem}\label{GWBChybrid} The exponents region $\mathcal{E}_{BC}^{\text{hyb}}$ is achievable, i.e., $$\mathcal{E}_{BC}^{\text{hyb}}\subseteq \mathcal{E}_{BC}^{\SI}.$$
\end{theorem}
\begin{IEEEproof} The region is achieved by the coding and testing scheme described in the previous subsection. This is proved in Appendix~\ref{app:noisy}.
\end{IEEEproof}
{To evaluate the region $\mathcal{E}_{BC}^{\text{hyb}}$, it suffices to consider auxiliaries whose alphabets satisfy the following two conditions: $|\mathcal{U}_0|\leq |\mathcal{X}|+8$, $|\mathcal{U}_1|\leq |\mathcal{X}| \cdot |\mathcal{U}_0|+3$ and $|\mathcal{U}_2|\leq |\mathcal{X}| \cdot |\mathcal{U}_0|+3$.} 

The exponents region $\mathcal{E}_{BC}^{\text{hyb}}$ is achieved by means of hybrid joint source-channel coding with side-information. The constraints in \eqref{thmhybrid} ensure that the receivers can decode their intended hybrid coding  codewords; a $U_0$-codeword  is decoded at both receivers and a $U_i$-codeword at Receiver $i$ only. These codewords are then used at the receivers for testing against conditional independence, see the exponents expression in \eqref{thmhybrid}. 
Notice that  hybrid joint source-channel coding also includes separate source-channel coding as a special case \cite{Minero}. In fact, the separate scheme's exponents region can be derived by considering $U_0 = (W_0,\tilde{U}_0)$ and $U_i =(W_i,\tilde{U}_i)$, for $i\in\{1,2\}$, where $(\tilde{U}_0,\tilde{U}_1,\tilde{U}_2,W_0,W_1,W_2)$ are auxiliary random variables which satisfy the Markov chains $(\tilde{U}_0,\tilde{U}_1,\tilde{U}_2)\to X\to (Z_1,Z_2)$ and $(W_0,W_1,W_2)\to W\to (V_1,V_2)$  and  the tuple $(W_0,W_1,W_2)$ is independent of $(\tilde{U}_0,\tilde{U}_1,\tilde{U}_1,X,Y_1,Z_1,Y_2,Z_2)$.

%

This theorem recovers the optimal error exponent for hypothesis testing over a point-to-point channel found in \cite{Gunduz}. It can be verified that the optimal error exponent of \cite{Gunduz} for the discrete memoryless channel from $W$ to $V_1$ can be  recovered by specializing Theorem \ref{GWBChybrid} to  $U_0, U_2$ constants and $U_1=(\tilde{U},W)$ with $W$ independent of $(\tilde{U},X,Y_1,Z_1)$. 


\subsection{An Example}

We investigate the achievable exponent region of Theorem~\ref{GWBChybrid} by means of an example. Reconsider the first example in  Section \ref{sec:example}, but where now communication takes place over a  Gaussian BC. Since the exponents region depends on the BC transition law only through the conditional marginals $P_{V_1|W}$ and $P_{V_2|W}$, we assume that the Gaussian BC is degraded and  described as follows:
\begin{align}
V_1 &= W+T_1,\\
V_2 &= V_1+T_2,
\end{align}
where $T_1$ and $T_2$ are independent Gaussian random variables of variances $r_1^2$ and $r_2^2-r_1^2$ ($r_2^2\geq r_1^2$). The input $W$ is subject to an expected power constraint $\mathbb{E}[|W|^2]\leq 1$. 

Likewise to  the first example, we  choose the auxiliaries 
 $U_0$ and $U_1$ jointly Gaussian with $X$  so that $X=U_1+Q_1$ and $U_1=U_0+Q_0$, and we choose $U_2=U_0$. Due to the degradedness of the channel, for such a choice of auxiliaries (i.e., when $U_2=U_0$) constraints \eqref{eq:conshyb}
 simplify to the two constraints 
 \begin{align}
 I(U_0;X) &\leq I(U_0;V_2),\\
 I(U_1;X|Z_1,U_0) &\leq I(U_1;V_1|Z_1,U_0).
 \end{align}

Let $Q_0, Q_1,U_0$ be independent zero-mean Gaussian random variables of variances $\sigma_{q_0}^2$, $\sigma_{q_1}^2$, and $1- \sigma_{q_0}^2-\sigma_{q_1}^2$ so that $X=Q_0+ Q_1+U_0$ and $U_1=U_0+Q_0$. Then, set the channel input to $W=\alpha U_0+\beta U_1$ for some parameters $\alpha, \beta \geq0$ satisfying 
\begin{equation}\label{eq:alphabeta}
(\alpha+\beta)^2 (1- \sigma_{q_0}^2-\sigma_{q_1}^2)+\beta^2\sigma_{q_0}^2=1.
\end{equation} 

Specializing the achievable exponents region $\mathcal{E}_{BC}^{\text{hyb}}$ to the proposed choices, proves achievability  of all nonnegative pairs $(\theta_1, \theta_2)$ that satisfy 
\begin{align}&\theta_1 \leq \nonumber\\&  \frac{1}{2}\log\left(\left( \sigma_1^2+\frac{\sigma_z^2}{1+\sigma_z^2}\right)\cdot \left( \frac{\sigma_{q_1}^2+\sigma_z^2}{\sigma_{q_1}^2(\sigma_1^2+\sigma_z^2)+\sigma_1^2\sigma_z^2} \right)\right),\\ 
\theta_2 &\leq \frac{1}{2}\log \left( \frac{1+\sigma_z^2+\sigma_2^2}{\sigma_{q_0}^2+\sigma_{q_1}^2+\sigma_z^2+\sigma_2^2} \right),\end{align}
for some $\sigma_{q_0}^2,\sigma_{q_1}^2\in[0,1]$, $\beta>0$ so that 
\begin{equation}\sigma_{q_0}^2 +\sigma_{q_1}^2 \leq 1,
\end{equation}
 and 
\begin{align}
\frac{1}{\sigma_{q_0}^2+\sigma_{q_1}^2} &\leq  \frac{1+r_2^2}{\beta^2\sigma_{q_0}^2+r_2^2} ,\label{g1}\\
\frac{1+\frac{\sigma_z^2}{\sigma_{q_1}^2}}{1+\frac{\sigma_z^2}{\sigma_{q_0}^2+\sigma_{q_1}^2}}  &\leq  1+\frac{\sigma_{q_0}^2}{\sigma_{q_1}^2+\sigma_z^2}+\frac{\beta^2 \sigma_{q_0}^2}{r_1^2}.\label{g2}
\end{align}

\begin{figure}[t]
	\centering
	\includegraphics[scale=0.3,angle=0]{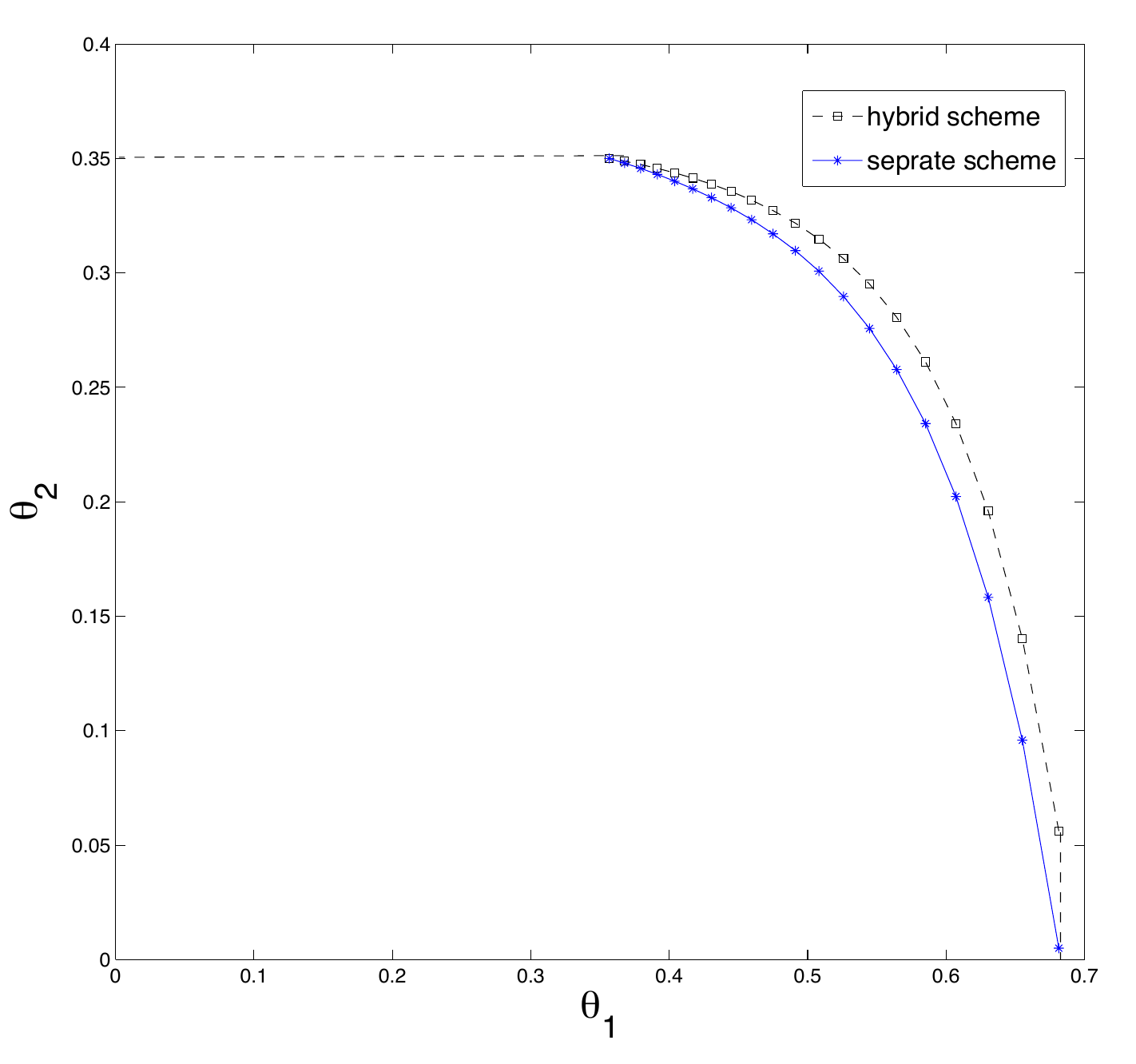}
	\caption{Achievable exponents region using hybrid scheme for $\sigma_z^2=0.7$, $\sigma_1^2=0.2$, $\sigma_2^2=0.3$, $r_1^2=0.1$, $r_2^2=0.3$.}
	\vspace{-0.8cm}
	\label{Figure 8}
\end{figure}
The boundary of the achievable exponents  region $\mathcal{E}_{BC}^{\text{hyb}}$  is illustrated in Fig.~\ref{Figure 8} for a  setup parametrized by $\sigma_z^2=0.7$, $\sigma_1^2=0.2$, $\sigma_2^2=0.3$, $r_1^2=0.1$ and $r_2^2=0.3$. One  observes a trade-off between the two exponents $\theta_1$ and $\theta_2$. Comparing this exponents region with the  region shown in Figure~\ref{Figure 5} for the noiseless channel, we observe that the asymmetric channel (different noise variances at the different receivers) changes the nature of this tradeoff. The second line shown in the figure depicts the boundary of the exponents region 
	that is achieved by a separation based scheme that combines the Gray-Wyner coordination coding with side-information from the previous section with a superposition code for the Gaussian broadcast channel. As it can be seen, the exponents region achieved by this separate coding and testing scheme is strictly smaller than the exponents region of our joint coding and testing scheme. 

\section{Conclusion and Discussion}\label{sec7}

This paper considers a distributed binary hypothesis testing problem in a one-observer, two-decision center setup. Achievable error exponents are presented for \emph{testing against conditional independence} when communication from the observer to the centers is over one common and two individual noise-free bit-pipes and when communication is over a BC. 
To this end, we presented coding and testing schemes where:
\begin{itemize} 
	\item  all terminals split their observations into many subblocks; 
	\item  transmitter and receivers apply a Gray-Wyner coordination code with  side-information \cite{Shayevitz} or hybrid joint source-channel coding with side-information for a BC;  
	\item the receivers apply a Neyman-Pearson test to the i.i.d. subblocks of side-information and reconstructed source sequences.
\end{itemize} 
Similarly to \cite{Wagner, Gunduz}, in the above approach, the ``multi-letter" decision over subblocks avoids introducing a competing error exponent due to the binning  or the channel decoding procedure.

The derived type-II error exponents are optimal when  \emph{testing against independence} over a common and two individual noise-free bit pipes, and when \emph{testing against conditional independence} over a single noise-free bit pipe if some of the  receiver side-informations are less noisy. 
An explicit characterization  of this latter optimal error exponent is given for a Gaussian example. This characterization clearly reveals a tradeoff between the error exponents achieved at the two decision centers. 

\section{Acknowledgement}
M. Wigger wishes to thank O. Shayevitz for helpful discussions.

\appendices
\section{Proof of Theorem~\ref{mainthm}}\label{app:1}

The proof is based on scheme in Section~\ref{sec:scheme2} which we analyze in the following.

\textbf{\textbf{Analysis}}:

From the way we constructed the Neyman-Pearson tests, it immediately follows that  the type-I error probabilities at the two receivers cannot exceed $\epsilon$. We turn our attention to  the type-II error probabilities. Notice that the analysis in \cite[Theorem 2]{Shayevitz} is easily modified to show  that for each $b\in\{1,\ldots, B\}$ and $i\in\{1,2\}$:
\begin{equation}
\Pr[ (X_b^k, \hat{U}_{i,b}^k, Z_{i,b}^k) \in \mathcal{T}_{\mu}^k (P_{U|X} P_{XZ} )] >1-\mu,
\end{equation}
for sufficiently large $k$. In fact, it suffices to add the sequence $Z_{i,b}^k$ into the typicality test defining event $\mathcal{E}_{3,i}$ in \cite[Appendix~B]{Shayevitz}. Thus, by the conditional typicality lemma \cite{ElGamal}, under the null-hypothesis  $\mathcal{H}=0$, also 
\begin{equation}\label{eq:mu}
\Pr[ (X_b^k,\hat{U}_{i,b}^k, Z_{i,b}^k,Y_{i,b}^k) \in \mathcal{T}_{\mu}^k (P_{U|X}P_{XYZ} )] >1-\mu.
\end{equation}

Now, recall that each Receiver~$i$ only declares $\hat{\mathcal{H}}_i=0$ if the applied Neyman-Pearson test produces $0$. 
Since  for each $i\in\{1,2\}$:
\begin{subequations}\label{eq:distributions}
\begin{IEEEeqnarray}{rCl}
&&\textnormal{ under }\mathcal{H}=0\colon\nonumber\\&&  \big\{\hat{U}_{i,b}^k, Y_{i,b}^k, Z_{i,b}^k\big\}_{b=1}^B\; \textnormal{ is i.i.d. } \nonumber\\&&\qquad\qquad\qquad\qquad\qquad\;\;\;\sim P_{\hat{U}_{i}^kY_i^kZ_i^k},
\end{IEEEeqnarray}
and 
\begin{IEEEeqnarray}{rCl}
&&\textnormal{ under }\mathcal{H}=1\colon \nonumber\\&& \big\{\hat{U}_{i,b}^k, Y_{i,b}^k, Z_{i,b}^k\big\}_{b=1}^B \;\textnormal{ is i.i.d. } \nonumber\\&&\qquad\qquad\qquad\qquad\;\;\;\sim P_{\hat{U}_{i}^kZ_i^k}P_{Y_i^k|Z_{i}^k},
 \end{IEEEeqnarray}
 \end{subequations}
 the Chernoff-Stein Lemma \cite{Cover} can be applied to bound the probabilities of type-II error. 
Thus, for sufficiently large $k$:
\begin{align}
- \frac{1}{n} \log\beta_{i,n} 
&\geq \frac{1}{k} D\big(P_{\hat{U}_i^kY_i^kZ_i^k|\mathcal{H}=0}\, \big\|\, P_{\hat{U}_i^kY_i^kZ_i^k|\mathcal{H}=1}\big)\nonumber\\&\qquad-\mu\nonumber\\
&\stackrel{(a)}{=}\frac{1}{k}  I\big(\hat{U}_i^k;Y_i^k\big|Z_i^k\big)-\mu\nonumber \\
&=    H\big(Y_i \big|Z_i)- \frac{1}{k} H\big(Y_i^k\big|\hat{U}_i^k, Z_i^k\big)-\mu,
\label{analynew23}
\end{align}
where mutual informations and entropies have to be computed according to the joint pmf $P_{\hat{U}_{i}^kY_i^kZ_i^k}$ under $\mathcal{H}=0$, and Equality $(a)$ holds by \eqref{eq:distributions}. 
We continue by defining the event $$\mathcal{E}_{V,i} \stackrel{\Delta}{=}\{(\hat{U}_i^k,Y_i^k,Z_i^k) \in \mathcal{T}_{\mu}^k(P_{{U}_iY_iZ_i})\},$$
and let  $\mathbbm{1}_V$ be the indicator function of $\mathcal{E}_{V,i}$.

The second term on the RHS of (\ref{analynew23}) can then be upper bounded as: 
\begin{align}
&H(Y_i^k|Z_i^k,\hat{U}_i^k) \nonumber\\&= H(Y_i^k,\mathbbm{1}_V|Z_i^k,\hat{U}_i^k)\nonumber \\
&= H(Y_i^k|Z_i^k,\hat{U}_i^k,\mathbbm{1}_V)+ H(\mathbbm{1}_V|Z_i^k,\hat{U}_i^k)\nonumber\\
&\stackrel{(a)}{\leq}H(Y_i^k|Z_i^k,\hat{U}_i^k,\mathbbm{1}_V)+1\nonumber\\
&\stackrel{(b)}{\leq} H(Y_i^k|Z_i^k,\hat{U}_i^k,\mathbbm{1}_V=1)+k\log|\mathcal{Y}_i|\cdot \mu+1\nonumber\\
&= \hspace{-.4cm}\sum_{\substack{({u}_i^k,z_i^k)\\\in \mathcal{T}_{\mu}^k(P_{U_iZ_i})}} \Big[\Pr[Z_i^k=z_i^k,\hat{U}_i^k={u}_i^k| \mathbbm{1}_V=1] \nonumber\\[-4ex]&\qquad\qquad\;\;\cdot  H(Y_i^k|Z_i^k=z_i^k,\hat{U}_i^k={u}_i^k,\mathbbm{1}_V=1) \Big] \nonumber \\
& \hspace{2cm}+k\log|\mathcal{Y}_i|\cdot\mu+1\nonumber\\
&\stackrel{(c)}{\leq} \hspace{-.4cm} \sum_{\substack{({u}_i^k,z_i^k)\\\in \mathcal{T}_{\mu}^k(P_{U_iZ_i})}} \Big[\Pr[Z_i^k=z_i^k,\hat{U}_i^k={u}_i^k|\mathbbm{1}_V=1] \nonumber\\[-4ex]&\qquad\qquad\qquad\qquad\cdot \log (|\mathcal{T}_{\mu}^k(Y_i^k|{u}_i^k,z_i^k)|)\Big] \nonumber\\& \qquad\qquad+k\log|\mathcal{Y}_i|\cdot\mu+1\nonumber\\
&\stackrel{(d)}{\leq} \hspace{-.4cm}\sum_{\substack{({u}_i^k,z_i^k)\\\in \mathcal{T}_{\mu}^kk(P_{U_iZ_i})}}  \Big[\Pr[Z_i^k=z_i^k,\hat{U}_i^k={u}_i^k|\mathbbm{1}_V=1]\nonumber\\[-5ex]&\qquad\qquad\qquad\cdot	(kH(Y_i|Z_i,{U}_i)+k\delta(\mu))\Big]\nonumber\\&\qquad\qquad\qquad +k \log|\mathcal{Y}_i|\cdot \mu+1\nonumber\\
&=kH(Y_i|Z_i,{U}_i)+k\delta(\mu)+ k\log|\mathcal{Y}_i|\cdot\mu+1. \label{eq:nextstep}
\end{align}
The steps leading to (\ref{eq:nextstep}) are justified as follows: 
\begin{itemize}
	\item
	$(a)$ follows from the fact  that $H(\mathbbm{1}_V|Z_1^k,\hat{U}_1^k)\leq 1$ because $\mathbbm{1}_V$ is a binary random variable;
	\item $(b)$ follows by \eqref{eq:mu}, because  $ \Pr[\mathbbm{1}_V=1]\leq 1$, and 
		because $H(Y_1^k|Z_1^k,\hat{U}_1^k,\mathbbm{1}_V=0)\leq k\log|\mathcal{Y}_1|$;
	\item $(c)$ follows because entropy is maximized by a  uniform distribution,
	\item $(d)$ follows by bounding the size of the typical set \cite{ElGamal} where $\delta(\mu)$ is a function that goes to 0 as $\mu\to 0$. 
\end{itemize}
We combine \eqref{analynew23} with \eqref{eq:nextstep} to obtain that for any choice of $\mu>0$ and sufficiently large $k, B$:
\begin{align}
	-\frac{1}{n}\log\beta_{i,n} 
	 \geq  I({U}_i;Y_i|Z_i) -\delta'(\mu), \qquad i\in\{1,2\},
\end{align}
where  $\delta'(\mu)$ is a function that tends to 0 as $\mu \to 0$. Taking $\mu\to 0$ proves  Theorem~\ref{mainthm}.


\section{Converse Proof to Theorem \ref{thm3}}\label{thm3p}

Fix a sequence of encoding and decoding functions $\{\phi^{(n)}, g_1^{(n)}, g_2^{(n)}\}$ so that the inequalities in Definition \ref{defHB} hold for sufficiently large blocklengths $n$. Fix also such a sufficiently large $n$. Then, define $U_{0,t}\stackrel{\Delta}{=}(M_0,Z_1^{t-1})$ and $U_{1,t}\stackrel{\Delta}{=}(X^{t-1},Z_{1,t+1}^n)$. {Following similar steps as in \cite{Michele}}, it can be shown that $$D(P_{M_0Y_1^nZ_1^n|\mathcal{H}=0}||P_{M_0Y_1^nZ_1^n|\mathcal{H}=1})
\geq -(1-\epsilon)\log \beta_{1,n}.$$ Therefore, the type-II error probability at Receiver~1 can be upper bounded as
\begin{align}
&-\frac{1}{n}\log \beta_{1,n}\nonumber\\&\leq \frac{1}{n(1-\epsilon)}D(P_{M_0Y_1^nZ_1^n|\mathcal{H}=0}||P_{M_0Y_1^nZ_1^n|\mathcal{H}=1})\nonumber\\
&\stackrel{(a)}{=}\frac{1}{n(1-\epsilon)}I(M_0;Y_1^n|Z_1^n)\nonumber\\
&=\frac{1}{n(1-\epsilon)}\sum_{t=1}^nI(M_0;Y_{1,t}|Y_1^{t-1},Z_1^n)\nonumber\\
&\stackrel{(b)}{\leq} \frac{1}{n(1-\epsilon)}\sum_{t=1}^nI(M_0,Y_1^{t-1},Z_1^{t-1},Z_{1,t+1}^n;Y_{1,t}|Z_{1,t})\nonumber\\
&\stackrel{(c)}{\leq} \frac{1}{n(1-\epsilon)}\sum_{t=1}^nI(M_0,X^{t-1},Z_1^{t-1},Z_{1,t+1}^n;Y_{1,t}|Z_{1,t})\nonumber\\
&=\frac{1}{n(1-\epsilon)}\sum_{t=1}^nI(U_{0,t},U_{1,t};Y_{1,t}|Z_{1,t}),\nonumber
\end{align}
where $(a)$ follows because under hypothesis $\mathcal{H}=1$ and given $Z_1^n$, the sequence $Y_1^n$ and message $M_0$ are independent;  $(b)$ follows from the memoryless property of the sources;  $(c)$ follows from the Markov chain $(Y_{1,t},Z_{1,t})\to (M_0,X^{t-1}, Z_1^{t-1}, Z_{1,t+1}^{n})\to Y_1^{t-1}$. 
For the type-II error probability at Receiver 2, one obtains: 
\begin{align}
&-\frac{1}{n}\log \beta_{2,n} \nonumber\\&\leq \frac{1}{n(1-\epsilon)}D(P_{M_0Y_2^n|\mathcal{H}=0}||P_{M_0Y_2^n|\mathcal{H}=1})\nonumber\\&=\frac{1}{n(1-\epsilon)}I(M_0;Y_2^n)\nonumber\\
&=\frac{1}{n(1-\epsilon)}\sum_{t=1}^nI(M_0;Y_{2,t}|Y_{2,t+1}^n)\nonumber\\
&=\frac{1}{n(1-\epsilon)}\sum_{t=1}^n\Big[ I(M_0,Z_1^{t-1};Y_{2,t}|Y_{2,t+1}^n)\nonumber\\&\qquad\qquad\qquad\qquad-I(Z_1^{t-1};Y_{2,t}|M_0,Y_{2,t+1}^n)\Big]\nonumber\\
&\stackrel{(b)}{=}\frac{1}{n(1-\epsilon)}\sum_{t=1}^n\Big[I(M_0,Z_1^{t-1},Y_{2,t+1}^n;Y_{2,t})\nonumber\\&\qquad\qquad\qquad\qquad-I(Z_1^{t-1};Y_{2,t}|M_0,Y_{2,t+1}^n)\Big]\nonumber\\
&\stackrel{(c)}{=}\frac{1}{n(1-\epsilon)}\sum_{t=1}^n\Big[I(M_0,Z_1^{t-1},Y_{2,t+1}^n;Y_{2,t})\nonumber\\&\qquad\qquad\qquad\qquad-I(Y_{2,t+1}^n;Z_{1,t}|M_0,Z_{1}^{t-1})\Big]\nonumber\\
&\stackrel{(d)}{\leq}\frac{1}{n(1-\epsilon)}\sum_{t=1}^n\Big[I(M_0,Z_1^{t-1},Y_{2,t+1}^n;Y_{2,t})\nonumber\\&\qquad\qquad\qquad\qquad-I(Y_{2,t+1}^n;Y_{2,t}|M_0,Z_{1}^{t-1})\Big]\nonumber\\
&=\frac{1}{n(1-\epsilon)}\sum_{t=1}^nI(M_0,Z_1^{t-1};Y_{2,t})\nonumber\\&=\frac{1}{n(1-\epsilon)}\sum_{t=1}^nI(U_{0,t};Y_{2,t}),\nonumber
\end{align}
where $(b)$ follows from the memoryless property of the sources; $(c)$ follows from Csiszar and K\"orner's sum identity \cite{ElGamal}; and  $(d)$ follows from the less noisy assumption and the Markov chain $(M_0,Y_{2,t+1}^n,Z_1^{t-1})\to X_t\to (Y_{1,t},Y_{2,t},Z_{1,t})$ which holds by the memoryless property of   the sources and because $M_0$ is a function of $X^n$. For the rate $R_0$, one finds:
\begin{align}
&nR_0\geq H(M_0)\geq I(M_0;X^n,Z_1^n)\nonumber\\
&=I(M_0;X^n|Z_1^n)+I(Z_1^n;M_0)\nonumber\\
&= \sum_{t=1}^n [I(M_0;X_t|X^{t-1},Z_1^n)+I(M_0;Z_{1,t}|Z_1^{t-1})]\nonumber\\
&=   \sum_{t=1}^n \Big[I(M_0,X^{t-1},Z_1^{t-1},Z_{1,t+1}^n;X_t|Z_{1,t})\nonumber\\&\qquad\qquad\qquad\qquad\qquad+I(M_0,Z_1^{t-1};Z_{1,t})\Big]\nonumber\\
&=   \sum_{t=1}^n \Big[I(X^{t-1},Z_{1,t+1}^n;X_t|M_0,Z_{1,t},Z_1^{t-1})\nonumber\\&\qquad+I(M_0,Z_1^{t-1};X_t|Z_{1,t})+I(M_0,Z_1^{t-1};Z_{1,t})\Big]\nonumber\\
&=   \sum_{t=1}^n \Big[I(X^{t-1},Z_{1,t+1}^n;X_t|M_0,Z_{1,t},Z_1^{t-1})\nonumber\\&\qquad\qquad\qquad\qquad\qquad+I(M_0,Z_1^{t-1};Z_{1,t},X_t)\Big]\nonumber\\
&\geq   \sum_{t=1}^n \Big[I(X^{t-1},Z_{1,t+1}^n;X_t|M_0,Z_{1,t},Z_1^{t-1})\nonumber\\&\qquad\qquad\qquad\qquad\qquad\qquad+I(M_0,Z_1^{t-1};X_t)\Big]\nonumber\\
&=   \sum_{t=1}^n [I(U_{1,t};X_t|Z_{1,t},U_{0,t})+I(U_{0,t};X_t)].\nonumber
\end{align}
Notice that by the memoryless property of  the sources and because $M_0$ is a function of $X^n$, the Markov chain $(M_0,Z_{1,t+1}^n,Z_1^{t-1},X^{t-1})\to X_t\to (Y_{1,t},Y_{2,t},Z_t)$ holds, and thus $(U_{0,t}, U_{1,t}) \to X_t \to (Y_{1,t}, Y_{2,t}, Z_t)$.
The proof is then concluded by combining these observations with standard time-sharing arguments which require introducing the auxiliary random  variables $T\in\{1,...,n\}$, $U_0\stackrel{\Delta}{=}(U_{0,T},T)$,  $U_{1}\stackrel{\Delta}{=}U_{1,T}$, $X\stackrel{\Delta}{=}X_T$, $Y_1\stackrel{\Delta}{=} Y_{1,T}$, $Y_2\stackrel{\Delta}{=} Y_{2,T}$, and $Z_{1} \stackrel{\Delta}{=} Z_{1,T}$.

\section{Evaluation of $\mathcal{E}_{\GW}^{\SI}(R_0,R_1=0,R_2=0)$ for the example in Section~\ref{sec:example}}\label{app:example}
 
 That the exponent pairs in \eqref{eq:subeqs} lie in $\mathcal{E}_{\GW}^{\SI}(R_0,R_1=0,R_2=0)$   can be seen by evaluating \eqref{thm1a} for  auxiliaries  $U_0$ and $U_1$ that are jointly Gaussian with $X$ and so that $X=U_1+W_1$ and $U_1=U_0+W_0$ for independent zero-mean Gaussians $W_1$, $W_0$ and $U_0$ that are of variances $\frac{\sigma_z^2}{(\sigma_z^2+1)2^{-2\tilde{\alpha}}-1}$, $(\sigma_z^2+1)2^{-2(\tilde{\alpha}+R_0)}-\sigma_z^2(1+\frac{1}{(\sigma_z^2+1)2^{-2\tilde{\alpha}}-1})$ and $(1+\sigma_z^2)(1-2^{-2(\tilde{\alpha}+R_0)})$, respectively.
 
That $\mathcal{E}_{\GW}^{\SI}(R_0,R_1=0,R_2=0)$   is no larger than the region in \eqref{eq:subeqs} is proved as follows. By the EPI:
 \begin{align}
 h(Y_2|U_0) &\geq \frac{1}{2} \log \big(2^{2h(Z_1|U_0)}+2^{2h(N_2)}\big), \nonumber \\
 h(Y_1|U_0,U_1,Z_1) &\geq \frac{1}{2} \log \big(2^{2h(X|U_0,U_1,Z_1)}+2^{2h(N_1)}\big).
 \end{align}
 Moreover, rate-constraint   on $R_0$ is equivalent to
 \begin{align}
R_0 &\geq  I(U_0;X)+I(U_1;X|U_0,Z_1)\nonumber\\&= h(X)-h(X|U_0)+h(X|U_0,Z_1)\nonumber\\&\qquad-h(X|U_0,U_1,Z_1)\nonumber\\
 &=h(X)-I(X;Z_1|U_0)-h(X|U_0,U_1,Z_1)\nonumber\\
 &=h(X)-h(Z_1|U_0)+h(Z_1|X,U_0)\nonumber\\&\qquad-h(X|U_0,U_1,Z_1)\IEEEeqnarraynumspace\nonumber\\
 &=h(X,Z_1)-h(Z_1|U_0)-h(X|U_0,U_1,Z_1),\label{gaus1}
 \end{align}
 where the last equality follows from the Markov chain $U_0\to X\to Z_1$.
 
 Defining  now
 \begin{align}
 \alpha & := h(X|U_0,U_1,Z_1) \quad \textnormal{and}\quad
 \beta  := h(Z_1|U_0),
 \end{align}
 above inequalities show that  $\mathcal{E}_{\GW}^{\SI}(R_0,R_1=0,R_2=0)$ is included in the set of all pairs $(\theta_1,\theta_2)$  that satisfy
 \begin{IEEEeqnarray}{rCl}
 	\theta_1 &\leq h(Y_1|Z_1) -  \frac{1}{2} \log \big(2^{2\alpha}+2^{2h(N_1)}\big),\label{eq:con} \\
 	\theta_2 & \leq h(Y_2) - \frac{1}{2} \log \big(2^{2\beta}+2^{2h(N_2)}\big),\label{eq:con1}	\end{IEEEeqnarray}
 for some choice of  parameters $\alpha \leq h(X|Z_1)$ and $\beta \leq h(Z_1)$ so that 
 \begin{equation}
 \label{eq:rate_cons}
 {(\alpha-h(X|Z_1))}+(\beta-h(Z_1)) \geq - R_0.
 \end{equation}
 Now, since  the right-hand sides of \eqref{eq:con} and \eqref{eq:con1}  are  decreasing in the parameters $\alpha$ and $\beta$, these parameters should be chosen so that   the rate-constraint~\eqref{eq:rate_cons} is satisfied with equality. In other words, for fixed $\alpha$, the optimal $\beta$ is obtained by solving \eqref{eq:rate_cons} under the equality constraint. Defining  $
 \tilde{\alpha}:=(\alpha - h(X|Z_1))\leq 0$ and expressing the optimal $\beta$ in terms of $\tilde{\alpha}$ then  establishes the desired inclusion of $\mathcal{E}_{\GW}^{\SI}(R_0,R_1=0,R_2=0)$
 in the set of  pairs $(\theta_1,\theta_2)$ given in \eqref{eq:subeqs}.

\section{Proof of Theorem~\ref{GWBChybrid}}\label{app:noisy}
We analyze the probability of error of the scheme in Section~\ref{sec:scheme4}.  
It immediately follows that  the type-I error probabilities at the two receivers cannot exceed $\epsilon$ from the way the Neyman-Pearson test is designed. Now, we consider  the type-II error probabilities. They can be upper bounded using the Chernoff-Stein lemma. 
Thus, for sufficiently large $k$:
\begin{align}
- \frac{1}{n} \log\beta_{i,n}
&\geq \frac{1}{k} D\big(P_{\hat{U}_i^kY_i^kZ_i^k|\mathcal{H}=0}\, \big\|\, P_{\hat{U}_i^kY_i^kZ_i^k|\mathcal{H}=1}\big)\nonumber\\&\qquad-\mu\nonumber\\
&\stackrel{(a)}{=}\frac{1}{k}  I\big(\hat{U}_i^k;Y_i^k\big|Z_i^k\big)-\mu\nonumber \\
&\geq  H\big(Y_i\big|Z_i\big)-\frac{1}{k}H\big(Y_i^k\big|Z_i^k,\hat{U}_i^k\big)-\mu,\nonumber
\label{analynew23bb}
\end{align}
where mutual informations and entropies have to be computed according to the joint pmf $P_{\hat{U}_{i}^kY_i^kZ_i^k}$ under $\mathcal{H}=0$, and Equality $(a)$ follows because under $\mathcal{H}=1$, the joint distribution of the variables decomposes as $P_{\hat{U}_{i}^kZ_i^k}P_{Y_i^k|Z_{i}^k}$. As shown in detail in \cite{Minero}, for sufficiently large values of $k$,  the rate constraints in \eqref{eq:rc1}--\eqref{rc8} ensure that   
\begin{align}
\Pr \big[ (\hat{U}_{i,b}^k,Y_{i,b}^k,Z_{i,b}^k)\in \mathcal{T}_{\mu}^k(P_{U_iY_iZ_i}) \big] > 1-\mu.
\end{align}

 Following similar steps as the ones leading to \eqref{eq:nextstep}, one obtains:
\begin{align}
H\big(Y_i^k\big|\hat{U}_i^k, Z_i^k\big) &\leq H(Y_i|Z_i,{U}_i)+ \log|\mathcal{Y}_i|\cdot\mu+\frac{1}{k}\nonumber\\&+\delta(\mu),
\end{align}
for a function $\delta(\mu)$ that tends to $0$ as $\mu\to 0$.
Thus, we get
\begin{IEEEeqnarray}{rCl}
	-\frac{1}{n}\log \beta_{i,n} 
	& \geq & I(U_i;Y_i|Z_i) -\log|\mathcal{Y}_i|\cdot\mu -\frac{1}{k}\nonumber\\&&-\delta(\mu). \IEEEeqnarraynumspace
\end{IEEEeqnarray}
Taking $\mu\to 0$ and $k\to\infty$ proves the theorem.

\begin{IEEEbiography} [{\includegraphics[width=1in,height=1.25in,clip,keepaspectratio]{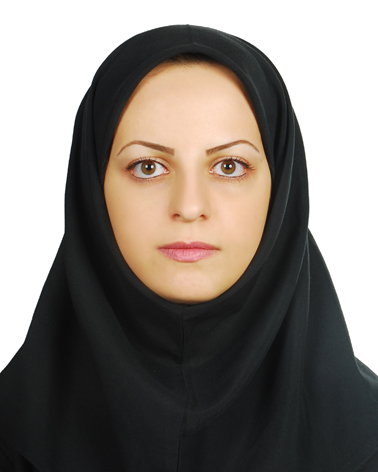}}]{Sadaf Salehkalaibar} (M'14) received the B.Sc., M.Sc. and Ph.D.  degrees in Electrical Engineering from Sharif University of Technology, Tehran, Iran in 2008, 2010 and 2014, respectively. She was a postdoctoral fellow at Telecom ParisTech, Paris, France in 2015 and 2017. She is currently an assistant professor at Electrical and Computer Engineering Department of University of Tehran, Tehran, Iran. Her special fields of interest include network information theory and fundamental limits of secure communication with emphasis on information-theoretic security.\end{IEEEbiography}

\begin{IEEEbiography} [{\includegraphics[width=1in,height=1.25in,clip,keepaspectratio]{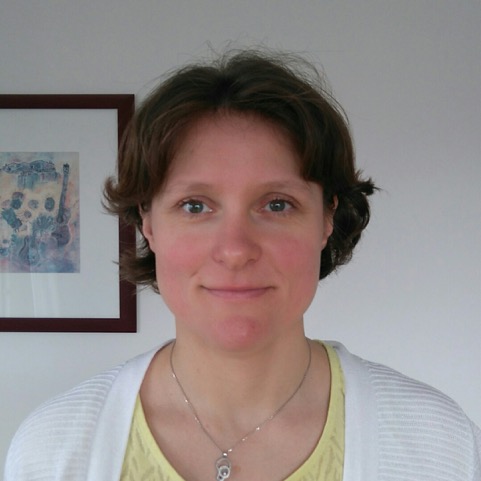}}]{Mich\`ele Wigger}   (S'05, M'09, SM'14) received the M.Sc. degree in electrical
	engineering, with distinction, and the Ph.D. degree in electrical engineering
	both from ETH Zurich in 2003 and 2008, respectively. In 2009, she was
	first a post-doctoral fellow at the University of California, San Diego, USA,
	and then joined Telecom Paris Tech, Paris, France, where she is currently an
	Associate Professor. Dr. Wigger has held visiting professor appointments at
	the Technion-Israel Institute of Technology and ETH Zurich. Dr. Wigger has
	previously served as an Associate Editor of the IEEE Communication Letters,
	and is now Associate Editor for Shannon Theory of the IEEE Transactions on
	Information Theory. She is currently also serving on the Board of Governors
	of the IEEE Information Theory Society. Dr. Wigger's research interests are in
	multi-terminal information theory, in particular in distributed source coding
	and in capacities of networks with states, feedback, user cooperation, or
	caching.\end{IEEEbiography}

\begin{IEEEbiography} [{\includegraphics[width=1in,height=1.25in,clip,keepaspectratio]{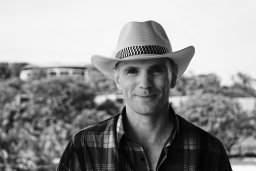}}]{Roy Timo}    is an Experienced Researcher at Ericsson Research in Stockholm,
	Sweden. Prior to joining Ericsson, he was an Alexander von Humboldt
	Research Fellow with the Institute for Communications Engineering at the
	Technische Universit¨at M¨unchen (2014-2016); a Research Fellow with the
	Institute for Telecommunications Research at the University of South Australia
	(2008-2013); and a Postdoctoral Researcher with the Department of Communications
	and Electronics at Telecom ParisTech (2013-2014). He received the
	Bachelor of Engineering and Ph.D. degrees from The Australian National
	University in 2005 and 2009, respectively.\end{IEEEbiography}


\begin{thebibliography}{30}
	\bibitem{Ahlswede} A. Ahlswede and I. Csiszar, ``Hypothesis testing with communication constraints,'' \emph{IEEE Trans. on Info. Theory}, vol. 32, no. 4, pp. 533--542, Jul. 1986.
	\bibitem{Han}
	T. S. Han, ``Hypothesis testing with multiterminal data compression,'' \emph{IEEE Trans. on Info. Theory}, vol. 33, no. 6, pp. 759--772, Nov. 1987.
	\bibitem{Amari}
	H. Shimokawa, T. Han and S. I. Amari, ``Error bound for hypothesis testing with data compression,'' in \emph{Proc. IEEE Int. Symp. on Info. Theory}, Jul. 1994, p. 114.
	\bibitem{Wagner}
	M. S. Rahman and A. B. Wagner, ``On the Optimality of binning for distributed hypothesis testing,'' \emph{IEEE Trans. on Info. Theory}, vol. 58, no. 10, pp. 6282--6303, Oct. 2012.
	\bibitem{Lai2}
	W. Zhao and L. Lai, ``Distributed testing against independence with conferencing encoders,'' in \emph{Prof. IEEE Inf. Theory Workshop (ITW)}, Korea, Oct. 2015.
	\bibitem{Kim}
	Y. Xiang and Y. H. Kim, ``Interactive hypothesis testing against independence,'' in \emph{Proc. IEEE Int. Symp. on Info. Theory}, Istanbul, Turkey, pp. 2840--2844, Jun. 2013.
	\bibitem{Debbah}
	G. Katz, P. Piantanida and M. Debbah, ``Collaborative distributed hypothesis testing,'' \emph{arXiv}, 1604.01292, Apr. 2016.
	\bibitem{Tan}
	J. Liao, L. Sankar, F. P. Calmon, V. Y. F. Tan, ``Hypothesis testing under maximal leakage privacy constraints'', To appear in \emph{Proc. IEEE Int. Symp. on Info. Theory}, Aachen, Germany, Jun. 2017.
	\bibitem{Gunduz}
	
	S. Sreekuma and D. Gunduz, ``Distributed hypothesis testing over noisy channels,'' available at: https://arxiv.org/abs/1704.01535.
	\bibitem{Shayevitz}
 	O. Shayevitz and M. Wigger, ``On the capacity of the discrete memoryless broadcast channel with feedback,'' \emph{IEEE Trans. on Inf. Theory}, vol. 59, no. 3, pp. 1329-1345, Mar. 2013.
	\bibitem{Gray}
	R. Gray and A. Wyner, ``Source coding for a simple network,'' \emph{Bell System Tech. J.}, vol. 48, pp. 1681--1721, Nov. 1974.
	\bibitem{Kaspi}
	A. Kaspi and T. Berger, ``Rate-distortion for correlated sources with partially separated encoders ,'' \emph{IEEE Trans. on Info. Theory}, vol. 28, no. 6, pp. 828--840, Nov. 1982.
	\bibitem{HBerger}
	C. Heegard and T. Berger, ``Rate distortion when side information may be absent,'' \emph{IEEE Trans. on Info. Theory}, vol. 31, no. 6, pp. 727--734, Nov. 1985.
	\bibitem{Minero}
	P. Minero, S. H. Lim, and Y. H. Kim, ``A unified approach to hybrid coding,'' \emph{IEEE Transactions on Information Theory}, vol. 61, no. 4, pp. 1509--1523, Apr. 2015.
	\bibitem{Cuff}
	P. W. Cuff and H. H. Permuter and T. M. Cover, ``Coordination capacity,''
	\emph{IEEE Trans. on Inf. Theory}, vol. 56, no. 9, pp. 4181-4206, Sept. 2010.
	\bibitem{ElGamal}
	A. El Gamal and Y. H. Kim, \emph{Network information theory}, Cambridge Univ. Press, 2011.
	\bibitem{Michele2}
	M. Wigger and R. Timo, ``Testing against independence with multiple decision centers,'' in \emph{Proc. of SPCOM 2016}, Bangalore, India, June 12-15, 2016. (Invited Paper)
	\bibitem{Csiszar1}
	I. Csiszar, ``Linear codes for sources and source networks: error exponents, universal coding,'' \emph{IEEE Trans. on Info. Theory}, vol. 28, no. 4, pp. 585--592, Jul. 1982.
	\bibitem{Wagner2}
	B. G. Kelly and A. B. Wagner, ``Improved source coding exponents via Witsenhausen's rate,'' \emph{IEEE Trans. on Info. Theory}, vol. 57, no. 9, pp. 5615--5633, Sep. 2011.
	\bibitem{Csiszar2}
	I. Csiszar and J. Korner, ``Graph decomposition: a new key to coding theorems,'' \emph{IEEE Trans. on Info. Theory}, vol. 27, no. 1, pp. 5--12, Jan. 1981.
	\bibitem{Tuncel2}
	E. Tuncel, ``Slepian Wolf coding over broadcast channels,'' \emph{IEEE Trans. on Info. Theory}, vol. 52, no. 4, pp. 1469--1482, Apr. 2006.
	\bibitem{Cover}
	T. M. Cover and J. A. Thomas, \emph{Elements of Information Theory}, John Wiley, 1991.
	\bibitem{Michele}
	S. Salehkalaibar, M. Wigger and L. Wang, ``Hypothesis testing in multi-hop networks'', available at: \url{https://arxiv.org/abs/1708.05198}.
	
	
	
	
	
	
	
	
	
	

	
	
	
	
	
\end{thebibliography}
\end{document}